\documentclass[apj,twocolumn]{openjournal}

\usepackage{amsmath,amssymb,amstext}
\usepackage[utf8]{inputenc}
\usepackage[breaklinks,colorlinks,
   urlcolor=blue,citecolor=blue,linkcolor=blue]{hyperref}
\usepackage{url}
\usepackage{lineno}
\usepackage{xcolor}
\usepackage{orcidlink}

\newcommand{\isotm}[2]{{}^{#2}\mathrm{#1}}

\graphicspath{ {./figures/} }

\journalinfo{The Open Journal of Astrophysics}

\begin{document}

\title{Multidimensional Nova Simulations with an Extended Buffer and Lower Initial Mixing Temperatures}
\author{
    Alexander Smith Clark$^1$ \orcidlink{0000-0001-5961-1680}
    Michael Zingale$^1$ \orcidlink{0000-0001-8401-030X}
}
\affiliation{$^1$Department of Physics and Astronomy, Stony Brook University, Stony Brook, NY 11794-3800, USA}
\begin{abstract}
A classical nova is a thermonuclear runaway initiated on a white dwarf accreting solar-like material from its stellar companion. Once the white dwarf accretes enough mass, the pressure at the base of the accreted layer reaches a critical point, leading to the ignition of the hydrogen fuel at their interface. This paper presents a set of two-dimensional CO classical nova simulations with an extended buffer zone of a fixed low density and temperature between the top of the accreted layer and the upper boundary, allowing us to capture the thermonuclear outburst in the domain.  Our domain reduces the role
of the upper-outflow boundary condition that has affected previous simulations and allows us to explore the nucleosynthesis evolution in detail.  We also study the effects of the initial temperature perturbation and buffer size to explore their sensitivity in our simulations. Finally, we start our simulations with a lower temperature at the base of the accreted layer ($7\times 10^7\,\mathrm{K}$) than previous work, allowing us to capture mixing earlier in the evolution, reducing the effects of the mixing-length-theory assumptions.  This allows for a more realistic description of convective transport in our models.
\end{abstract}

\maketitle

\section{Introduction}

A classical nova explosion begins as a white dwarf (WD) that accretes solar-like material from its companion star. 
As more matter settles onto the white dwarf, the core envelope interface (CEI) between the solar-like accreted material and the white dwarf becomes subject to immense pressure and electron degeneracy, igniting a thermonuclear runaway (TNR) that burns (a fraction of) the accreted mass, ejecting $10^{-5}$ to $10^{-3}\,M_{\odot}$ \citep{sparks:1978, livio:1994, gehrz:1998, jordi:2024}. Observations show (see, e.g., 
\citealt{livio:1990,livio:1994,gehrz:1998}), that there is an enrichment of the novae ejecta in $\isotm{C}{}$, $\isotm{N}{}$, $\isotm{O}{}$, and $\isotm{Ne}{}$ of $\gtrsim 30\%$ by mass. 
This suggests a mixing mechanism that dredges material from the white dwarf into the solar-rich layers.  Understanding this ``dredge-up'' has been a long-standing problem in
nova theory and simulations.

\citet{shara:1981, shara:1982} introduced one of the first dredge-up mechanisms, relying on the existence of local eruptions due to the long accreted layer thermalization timescales relative to the thermonuclear runaway timescales. However, neglecting convective heat transport and the azimuthal direction of the heat flow leads to several inconsistencies, as pointed out in \cite{fryxell:1982,orio:1993,shankar:1992,glasner:1995}.
Further mechanisms have been proposed to explain the convective energy transport and turbulent mixing effects: diffusive-mixing \citep{prialnik:1984, prialnik:1985, iben:1985, iben:1992}, mixing by viscous shear-instabilities \citep{kippenhahn:1978, mcdonald:1983, kutter:1987, fujimoto:1988, kutter:1989}, and convective overshooting \citep{glasner:1995, kercek:1998, kercek:1999}. 
These mechanisms are reviewed in \cite{livio:1990}.
Many of these proposed mechanisms, within the context of 1D simulations, were based on two important assumptions: a) the assumption of a diffusion transport model for species under turbulent-mixing \citep{cloutman:1976, prialnik:1979,brandenburg:2009}, and
b) the use of mixing-length-theory (MLT) and simplified linear analysis to construct a convective-transport model (see, for example, \citealt{kutter:1972, kutter:1980}). 

Ultimately, one-dimensional models are insufficient to capture
the physics needed to describe the mixing.
Turbulent-diffusive mixing requires a turbulent regime that is only valid when convection has already become the main heat transport mechanism from the WD to the accreted layer. 
The MLT assumptions require a hydrodynamic timescale far below the sound-speed propagation timescales \citep{glasner:1997}. Hence, the use of MLT is not valid near the TNR explosion, where the radial velocity Mach number is $\mathcal{O}(1)$. 
Finally, the azimuthal direction plays a significant role in the total heat transport of the convective layers and should not be ignored, as pointed out by \cite{orio:1993}. 

The first multidimensional simulations to address the shortcomings of the one-dimensional models were performed by \cite{shankar:1992, shankar:1994}, \cite{glasner:1995} and \cite{glasner:1997}. 
These simulations were constructed by mapping the initial Lagrangian 1D models, under the assumption of MLT in regions of convective instabilities, to a 2D-grid that can operate under a combination of Eulerian and Lagrangian formulations through a flexible moving mesh.  They demonstrated the need to use multidimensional simulations to accurately capture  the onset of the nova outburst.

In \cite{kercek:1998}, 2D simulations of CO-novae outbursts were explored in low- and high-resolution with the same initial models as in \cite{glasner:1997}. These simulations used an Eulerian code, with a Cartesian grid and periodic lateral boundary conditions.
In \cite{kercek:1999}, a 3D extension was proposed for two different initial models, consistent with the one presented in \cite{glasner:1997} with two accreted layer metallicities: a solar metalicity  ($\mathrm{Z}=0.02$) and an 
enriched metallicity ($\mathrm{Z}=0.1$). 
The results from \cite{kercek:1998, kercek:1999} suggested a less violent TNR than found in \citet{glasner:1997}, with lower velocities and temperature at the end of their simulations.
The differences between the \cite{kercek:1998, kercek:1999} and \cite{glasner:1997} motivated an important discussion about the sensitivity of the outer boundary conditions and the difference between the Eulerian (\texttt{PROMETHEUS}, \citealt{prometheus}) and Lagrangian (\texttt{VULCAN}, \citealt{livne:1993}) codes used in their studies. 
\citet{glasner:2005} compared the results of a Lagrangian and an Eulerian code with free, closed, and inflow-outflow balanced boundary conditions. In simulations where matter was allow to flow out through the upper boundary, they observed a drop in temperature (quenching) at the CEI, which they suggest is responsible for the TNR differences seen in the earlier studies.

\citet{glasner:2007} explored a one-dimensional  
initial model that becomes convective at a temperature of $3\times 10^{7}\,\mathrm{K}$ and evolved it using MLT without chemical mixing or diffusive-energy transport effects \citep{eggleton:1971} to create snapshots
with different CEI temperatures: $T_\mathrm{CEI} = 3.0\times 10^7\,\mathrm{K}$ (T3), $5.0\times 10^7\,\mathrm{K}$ (T5), $7.0\times 10^7\,\mathrm{K}$ (T7), 
$9.0\times 10^7 \,\mathrm{K}$ (T9), and $1.0\times 10^8 \,\mathrm{K}$ (T10).
Further multidimensional simulations were implemented using this model with both a Lagrangian and Eulerian code, with a typical resolution of $1.4\,\mathrm{km}\times 1.4\,\mathrm{km}$. 
This work shows a universal evolution of the nova, regardless of the starting initial model and independent of their initial perturbation size, nature, and intensity. 

More recently, \citet{casanova:2010} presented a set of runs using the initial model from \citet{glasner:1997, kercek:1998} with hydrostatic boundary conditions at the top and bottom (based on \citealt{zingale:2002}) with the velocity reflected at the bottom and outflow at the top. Several further studies explored the role of the Kelvin-Helmholtz instabilities as the main source of convective mixing \citep{casanova:2011a, casanova:2011b}, the role of the mass and the WD composition \citep{casanova:2016,casanova:2018}, and the transition between one-dimensional and three-dimensional models \citep{casanova:2020}. A summary of these simulations can be found in \citet{jordi:2024}. All these two- and three-dimensional runs start from initial  models with a CEI temperature of $10^{8}\,\mathrm{K}$, while matching the upper domain boundary with the top of the accreted layer, enforcing the HSE there. 

The goal of the present work is to explore these assumptions through a series of two-dimensional Eulerian simulations of the nova problem.
We start with a lower CEI temperature than previous works. Although the universality arguments provided by \cite{glasner:2007} suggest a small influence of the initial model CEI temperature choice in the evolution towards the TNR, by starting our runs with $T_\mathrm{CEI}\sim 7\times 10^{7}\,\mathrm{K}$ instead of $\sim 10^{8}\,\mathrm{K}$, we reduce the artificial contributions of MLT in our calculations. Additionally, temperatures close to $\sim 10^{8}\,\mathrm{K}$ are the threshold for $\beta^{+}$-decay channels of the cold-CNO cycles to freeze compared  to the fast $p$-capture enhanced channels of the hot-CNO cycle.
This
requires more nuclei and $\beta^{+}$-decay rates in the reaction network, which may have a large impact on the overall evolution of its nucleosynthesis. In earlier papers, \cite{casanova:2010,casanova:2011a,casanova:2011b, casanova:2016,casanova:2018} presented a network of 13 nuclei connected by 18 reactions. In our work we present a total of 17 nuclei connected by 31 reactions, including more $\beta^{+}$ decays and $p$-captures, which will give a more accurate representation of the energy generation.

We will also include a buffer region at the top boundary instead of matching the top of the accreted layer to the upper boundary. The purpose of this buffer is to extend the domain, allowing enough space for the TNR to occur and expand without losing mass through the upper boundary. This reduces the artificial quenching effects that the open-boundary case generates \citep{glasner:2005}. Finally, we also extend the CO-layer inward, toward the lower boundary, by assuming it to be isothermal, reducing
the influence of the lower boundary condition.
These changes allow our simulations to capture the maximum energy generation peak of the TNR inside the domain, maintaining a more accurate convection mixing description from an earlier stage in the evolution. Finally, we perform a sensitivity study on the buffer size and the resolution to compare to the results of 
\citet{casanova:2010, casanova:2011a, casanova:2011b, casanova:2018}.

\section{Initial Model}

We start with the T7 initial model of \citet{glasner:2007}---this was constructed with a $1.14 \,\mathrm{M}_{\odot}$ CO-white dwarf cooled to the point where the luminosity is about $1.6\times 10^{-2}\,L_{\odot}$, followed by  accretion solar-like material ($Z=0.02$) at a rate of 
$10^{-9}\,\mathrm{M}_{\odot}\,\mathrm{yr}^{-1}$. Once the base of the accreted layer reaches a temperature of $\sim 3 \times 10^{7}\,\mathrm{K}$, it becomes unstable to convection, and MLT is used to continue the evolution until the base temperature of the accreted layer reaches $T_{\mathrm{CEI}}\sim 7 \times 10^{7}\,\mathrm{K}$. The 1D profile consists of a 25 km portion of the CO WD followed by the accreted layer, encompassing $341\,\mathrm{km}$. The reaction network used for the model contained $15$ nuclei: $\isotm{H}{1}$, $\isotm{He}{3,4}$, $\isotm{Be}{7}$, $\isotm{B}{8}$, $\isotm{C}{12,13}$, $\isotm{N}{13\mbox{--}15}$, $\isotm{O}{14\mbox{--}17}$, and $\isotm{F}{17}$. Since this
is a Lagrangian model, the grid cells are not equally spaced in radius,
and the introduction of MLT generates a set of non-zero convective velocities, breaking the hydrostatic equilibrium assumption.  This requires an interpolation
procedure for our code.

\begin{figure}[t]
\centering
\epsscale{1.10}
\plotone{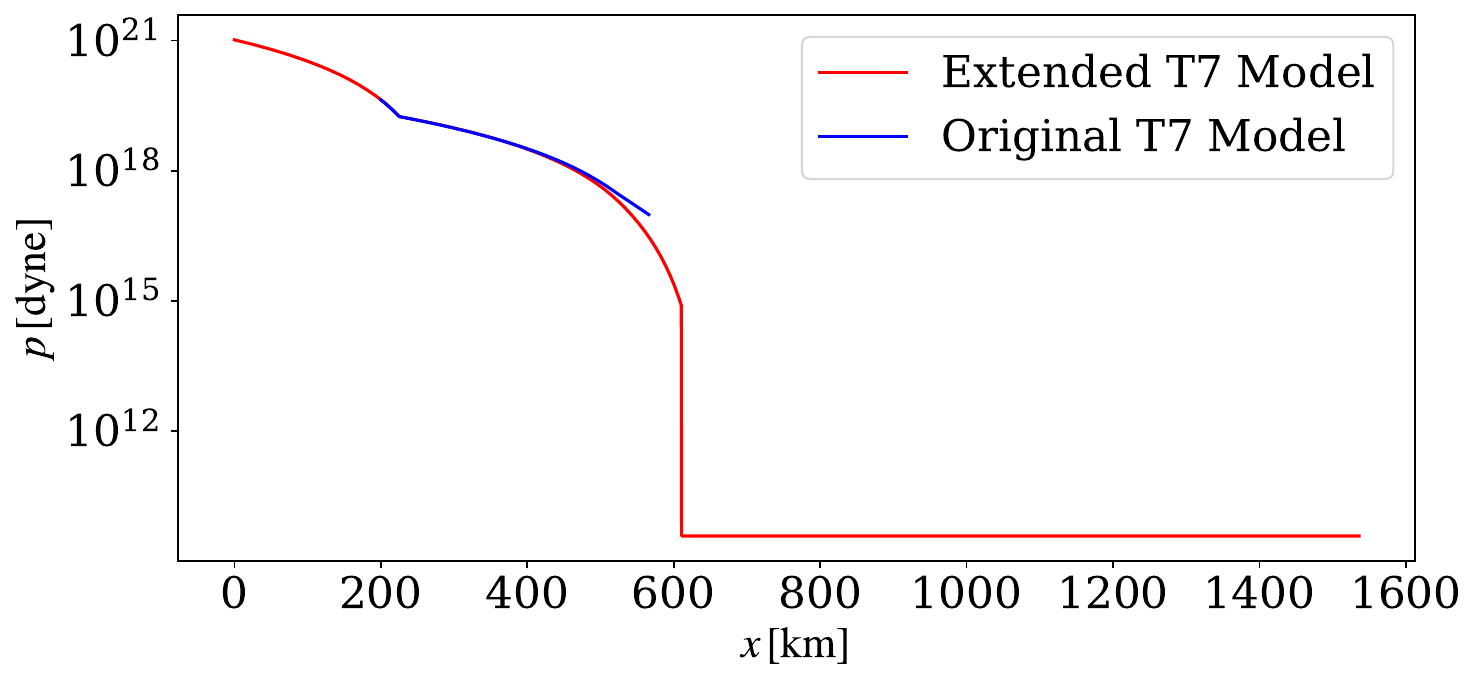} \\
\plotone{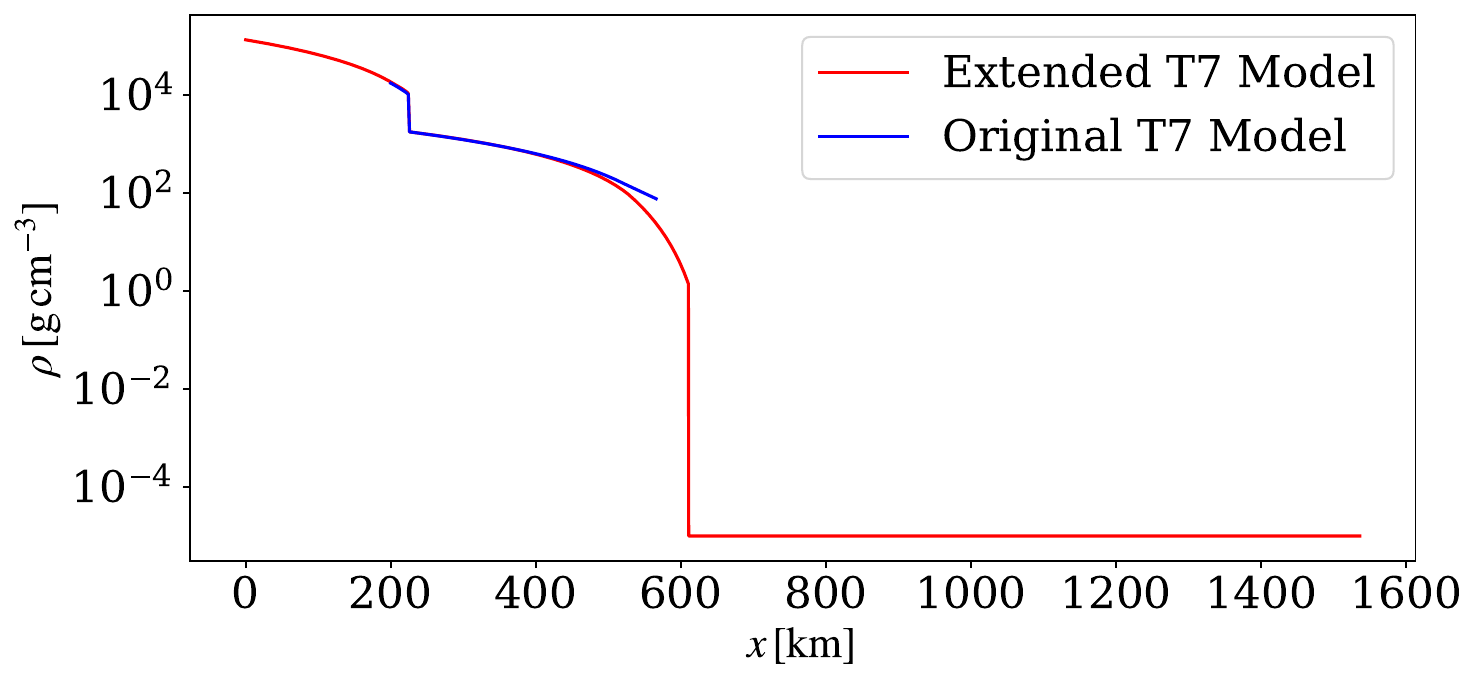} \\
\plotone{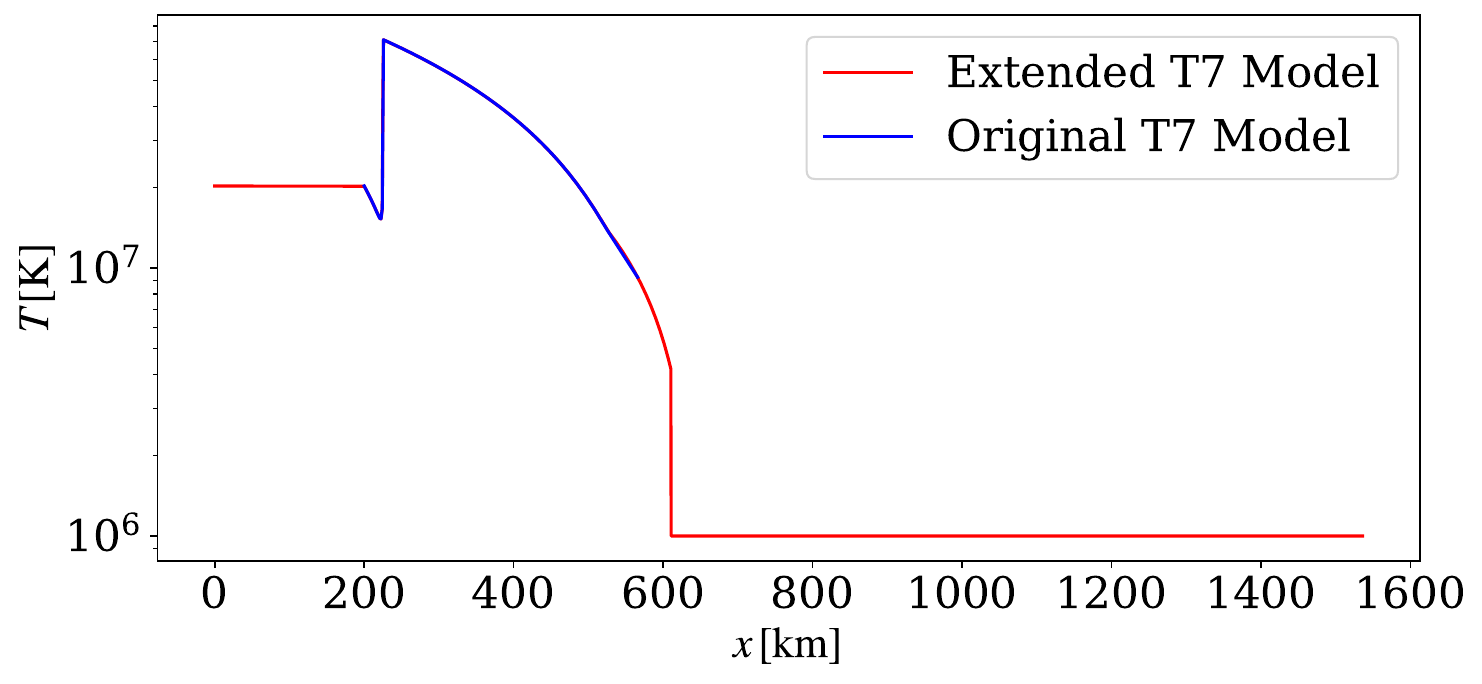} \\
\plotone{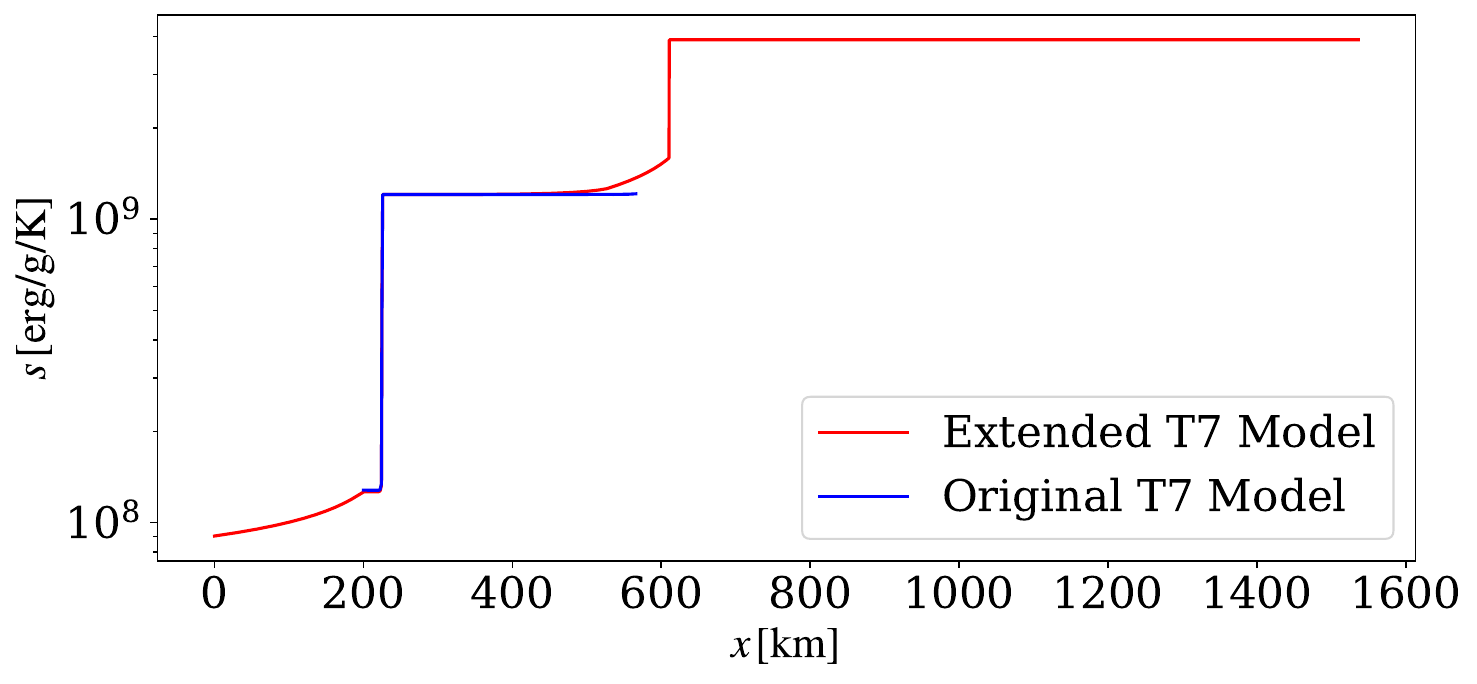}
\caption{ A comparison between the original T7 initial model thermodynamic variables profile described in \citep{glasner:2007} and, from our model A, on a uniform grid enforced to be thermodynamically consistent and in HSE. Note that the discrepancy between the density and pressure of the original T7 model and ours is originated by the fact that the original T7 initial model uses point gravity and has non-zero velocities.}
\label{fig:inital_models}
\end{figure}

To prepare the T7 model for mapping onto our domain, we first interpolate
it onto a uniform grid.  As part of this, we extend
the region beneath the CEI by $200~\mathrm{km}$ by assuming an isothermal temperature profile.  We then
integrate HSE starting at the CEI, following the procedure of 
\citet{zingale:2002}, and enforcing thermodynamic consistency with our EOS \citep{timmes_swesty:2000}, under the assumption of constant 
gravity\footnote{This value of $g$ was computed based on a radius $r=4.64\times 10^8\,\mathrm{cm}$ located in the original initial model domain.}, $g =-7.06\times 10^{8}\,\mathrm{cm/s^2}$.   

The HSE integration begins with locating the position of the CEI, based
on the radius of the maximum temperature.  We integrate the HSE condition both
inward and outward from this point. As we integrate upwards, we use the zone $i-1$ to find the state in zone $i$ using a second-order accurate discretization of HSE:
\begin{equation}
        \dfrac{P_i - P_{i-1}}{\Delta r} = \dfrac{1}{2}(\rho_{i} + \rho_{i-1})g
\end{equation}
   or after relabeling $P_i\rightarrow P_i^{\mathrm{HSE}}$
\begin{equation}
        P_{i,\mathrm{HSE}}= P_{i-1}+ \dfrac{\Delta r}{2}(\rho_{i} + \rho_{i-1})g \label{eq:hse}
\end{equation}
where $g$ is constant and $\Delta r$ is the uniform radial grid spacing. As we integrate downwards, we proceed similarly to the previous case, but instead with a forward finite difference:
\begin{equation}\label{eq:hse_d}
        P_{i,\mathrm{HSE}} = P_{i+1} - \dfrac{\Delta r}{2}(\rho_{i+1} + \rho_{i})g 
\end{equation}
The core idea of this interpolation method is to compute the pressure from (\ref{eq:hse}) and the equation of state by enforcing: 
    \begin{equation}
        P_{i,\mathrm{EOS}}(\rho_{i,\star}) - P_{i,\mathrm{HSE}}(\rho_{i,\star}) = 0
    \end{equation}
    to finally compute $\rho_{i,\star}$ and its respective pressure $P_{i,\star}$. Following the Newton-Raphson root-finding method over $\delta \rho^{(\nu)}_i$:
    \begin{equation}
        \delta \rho^{(\nu)}_i = \rho^{(\nu +1)}_i - \rho^{(\nu)}_i 
    \end{equation}
over each $\nu$-iteration:
    \begin{equation}\label{eq:delta_rho}
        \delta \rho^{(\nu)}_i = -\dfrac{P_{i,\mathrm{EOS}}^\nu - P_{i,\mathrm{HSE}}^\nu}{\left(\dfrac{dP}{d\rho}\right)_{i,\mathrm{EOS}}^\nu - \left(\dfrac{dP}{d\rho}\right)_{i,\mathrm{HSE}}^\nu}  
    \end{equation}
    where the $\nu$-superscript represents a quantity evaluated in $\rho^{(\nu)}$. 
All of the needed derivatives are obtained from the equation of state.  And we iterate on
each zone until $|\delta_i^{(\nu)}|$ is small.

Finally, we place a buffer-zone on the top of the accreted layer with a density of $10^{-5}\,\mathrm{g}\,\mathrm{cm}^{-3}$ and a temperature of $10^{6}\,\mathrm{K}$, filling the remainder of the 1D domain.
We construct two different initial models, differing only in the size 
of the buffer region, with model A having a vertical extent of $1536\,\mathrm{km}$ and model B an extent of $1024\,\mathrm{km}$.
The interpolated model is illustrated in Figure~\ref{fig:inital_models}. The specific entropy profile in Figure~\ref{fig:inital_models} was reconstructed by using the EOS \citep{timmes_swesty:2000}.  It
clearly shows that the accreted layer will be convectively
unstable.

\section{Numerical Method}

\begin{figure*}[t]
\centering
\epsscale{0.8}
\plotone{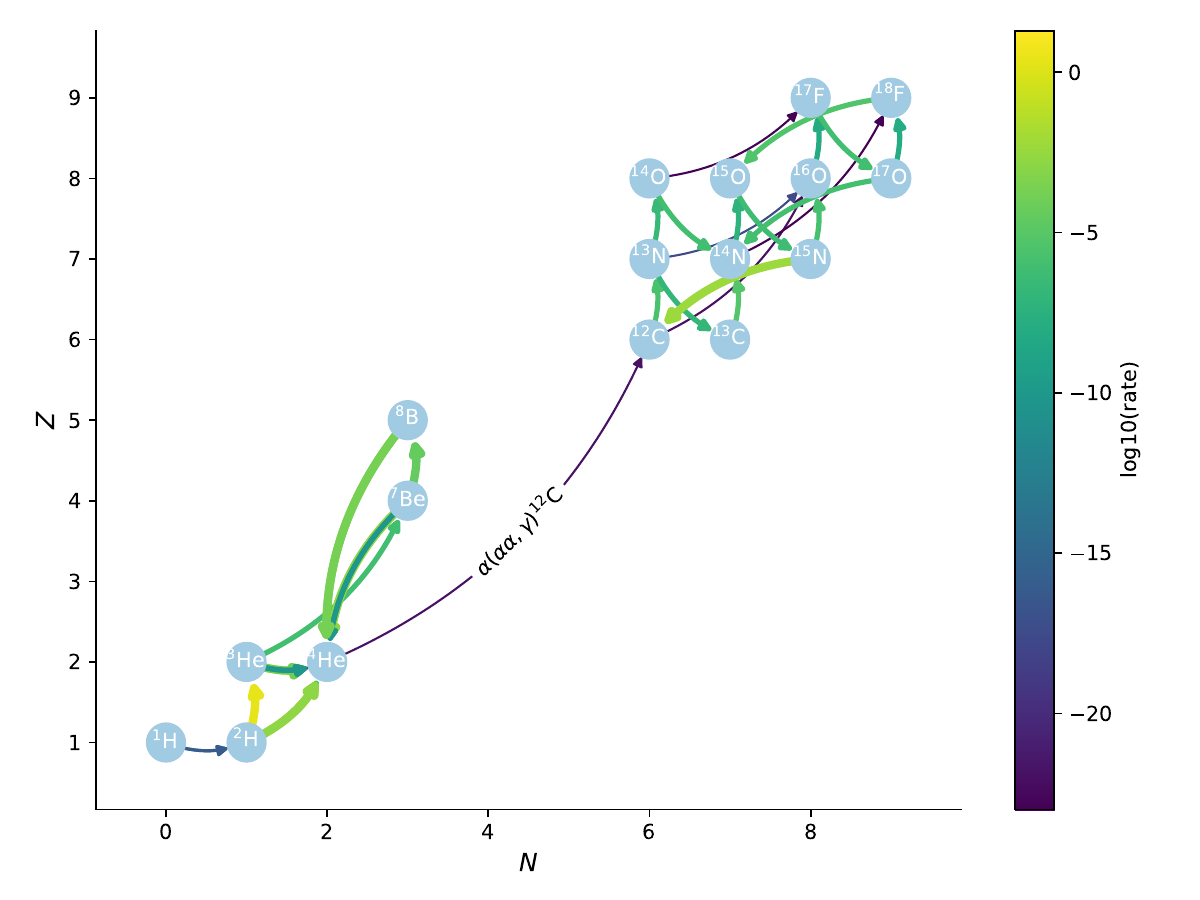}
\caption{An illustration of the reaction network with the rates evaluated
at  $\rho=1.0\times 10^{4}\,\mathrm{g} \,\mathrm{cm}^{-3}$ and $T=7.0\times 10^{7}\,\mathrm{K}$. At these conditions, $T\lesssim 10^8\,\mathrm{K}$, so the $p$-captures and $\beta^{+}$-decays compete over the $\beta^{+}$-unstable nuclei, like $\isotm{N}{13}$ and the CNO cycle is limited by the slowest $p$-capture reaction: $\isotm{N}{14}(p,\gamma)\isotm{O}{15}$.}
\label{fig:reaction}
\end{figure*}

We use \texttt{CASTRO}, an open-source compressible astrophysical simulation code \citep{castro,castro_joss}.  \texttt{CASTRO} is built on the \texttt{AMReX} adaptive mesh refinement library \citep{amrex_joss}, enabling it to focus resolution in regions of complex flow and to take advantage of GPU-based supercomputers using performance portable software abstractions \citep{castro-gpu}.  Our simulations use the corner-transport-upwind (CTU) numerical scheme \citep{colella:1990} with piecewise parabolic reconstruction and characteristic tracing \citep{ppm,millercolella:2002} to evolve  the conserved fluid state. \texttt{CASTRO} also employs the dual energy formalism described in \cite{bryan:1995}. The Riemann solver used to compute the interface state from the left and right interface reconstruction is described in \cite{castro} based on the ideas of \cite{bell:1989}. For the present simulations, we have included thermal diffusion in the energy equation, treated in an explicit-in-time fashion as described in \citet{eiden:2020}. 

As described earlier, considerable attention has been paid to the treatment of the upper
boundary condition, especially with Eulerian codes.  For these simulations, we place
a buffer region of low density material between the top of the atmosphere and the upper
boundary and use a zero-gradient (outflow), with the ghost cells initialized with the ambient state corresponding to the buffer region.  Furthermore, a sponging term is used in the buffer, based on that in \cite{wdmergerI} to prevent large velocities from building up.  These features allow the atmosphere to expand in response to the energy release without mass
flowing out of the upper boundary.  
We use a simple well-balanced approach with PPM to maintain hydrostatic equilibrium, based on the ideas in \cite{kappeli:2016,hse-rnaas}---this allows us to use a reflecting lower boundary condition for the domain.  Finally, the lateral boundary conditions are periodic.

Reactions are coupled to the hydrodynamics via Strang-splitting \citep{strang:1968} with energy
and mass fractions evolved together using the VODE integrator \citep{vode}.  This ensures that the overall method second-order
accurate in time \citep{strang_rnaas}.  Our reaction network consists of 17 nuclei: $\isotm{H}{1,2}$, $\isotm{He}{3,4}$, $\isotm{Be}{7}$, $\isotm{B}{8}$, $\isotm{C}{12,13}$, $\isotm{N}{13\mbox{--}15}$, $\isotm{O}{14\mbox{--}17}$, $\isotm{F}{17,18}$ (see Fig.~\ref{fig:reaction}) and is generated using  \texttt{pynucastro} \citep{pynucastro2} with the reaction rates from the \texttt{REACLIB} nuclear data library \citep{cyburt:2010}. To close the hydrodynamic system of equations, we use the Helmholtz equation of state \citep{timmes:1999, timmes_swesty:2000} and the stellar
conductivities of \citet{timmes:2000}.

Convection is seeded by introducing a temperature perturbation over the initial temperature profile, $T_0$, of the form:
\begin{eqnarray}
    \delta T &=& T_o\times \Delta\left[1+\cos\left(\dfrac{10\pi x}{L_x}\right) \right]\mathrm{exp}\left[-\left(\dfrac{y_L}{\sigma}\right)^2\right]\\
    y_L &=& y - (y_{CEI} + \alpha\sigma) 
\end{eqnarray}
where $\Delta$ is the perturbation maximum amplitude, $L_x$ is the horizontal length of the domain, $\sigma$ is the width of the perturbation, and $\alpha$ is the width-ratio factor, where $\alpha = 0$ sets the maximum amplitude location at the CEI, and $\alpha=1$ fixes the maximum amplitude location at $\sigma$ above the the CEI. For our models, we use $\alpha=1.8$ and $\sigma=10\,\mathrm{km}$.

We create two sets of simulations.  The ``A''-series simulations use a domain with a size ($D_x \times D_y$) of $3072~\mathrm{km} \times 1536~\mathrm{km}$ and a maximum resolution of $\Delta x_\mathrm{max} = 0.4\,\mathrm{km}$ (model A4) or $0.8\, \mathrm{km}$ (model A8), respectively, and a maximum initial perturbation size of $5\%$ of the initial temperature value ($\Delta = 0.05$).  Both models use a base grid ($N_x \times N_y$) of $1920\times 960$ cells with one level of refinement, with the A4 model using a $4\times$ jump and the A8 model using a $2\times$ jump in refinement.  The adaptive mesh refinement algorithm tags cells for refinement with the following criteria:
\begin{eqnarray}
    \rho &>& 10^2\, \mathrm{g}\, \mathrm{cm}^{-3}\\ 
    \dot{e}_{\mathrm{nuc}} &>&10^{11}\,\mathrm{erg}\,\mathrm{g}^{-1}\,\mathrm{s}^{-1}\\
    T&>&3\times 10^7\,\mathrm{K}
\end{eqnarray}
For the ``B''-series of simulations, we explore resolutions of $0.8\,\mathrm{km}$ (model B8), $0.4\, \mathrm{km}$ (model B4), and $0.2\,\mathrm{km}$ (model B2) respectively. The domain size of each model B is $2048~\mathrm{km} \times 1024~\mathrm{km}$. The B8 simulation uses a coarse grid of $1280\times640$ cells, with a single refinement level with a jump of $2\times$.   For the B4 model, we use a coarse grid of $640\times 320$ cells, with three levels of refinement, each a jump of $2\times$.  Finally,
for the B2 model, we again use a coarse grid of $1280\times 640$ cells two
two levels of refinement, the first a jump of $4\times$ and the next a jump of $2\times$.  The refinement criteria on density and energy generation rate
are the same as the ``A''-series models, and the temperature criteria is:
\begin{equation}
    T >10^7\,\mathrm{K}
\end{equation}
The use of adaptive mesh refinement in our models guarantees that only the accreted layer is continuously refined as the convective front moves forward in the domain, leaving the buffer region at coarser resolution.  As the envelope expands in time, more of the domain becomes refined.  This helps reduce the computational resources of the simulations.

All models except B8 were run on NVIDIA A100 GPUs on the ALCF Polaris machine.  B8
used CPUs (MPI + OpenMP) on the NERSC Perlmutter machine.
All of the simulation code necessary to run these simulations is available in our github repositories.
Table~\ref{table:models} summarizes the simulation setups and includes the time 
when material hits the top boundary (our stopping time), $t_\gamma$, and the final
temperature at the CEI, $T_\mathrm{CEI}$.
\begin{figure*}[t]
\centering
\epsscale{2.1}
\plottwo{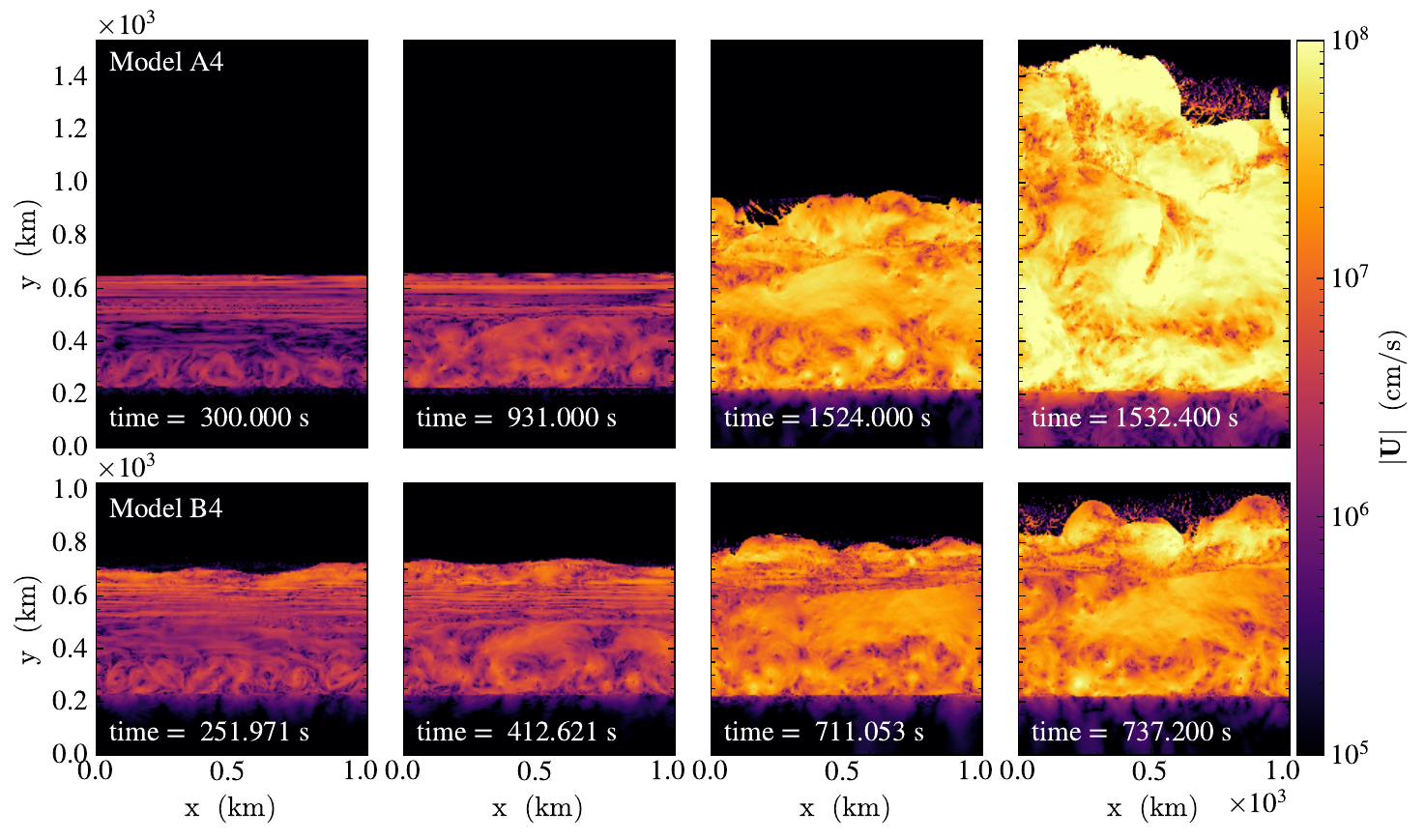}
{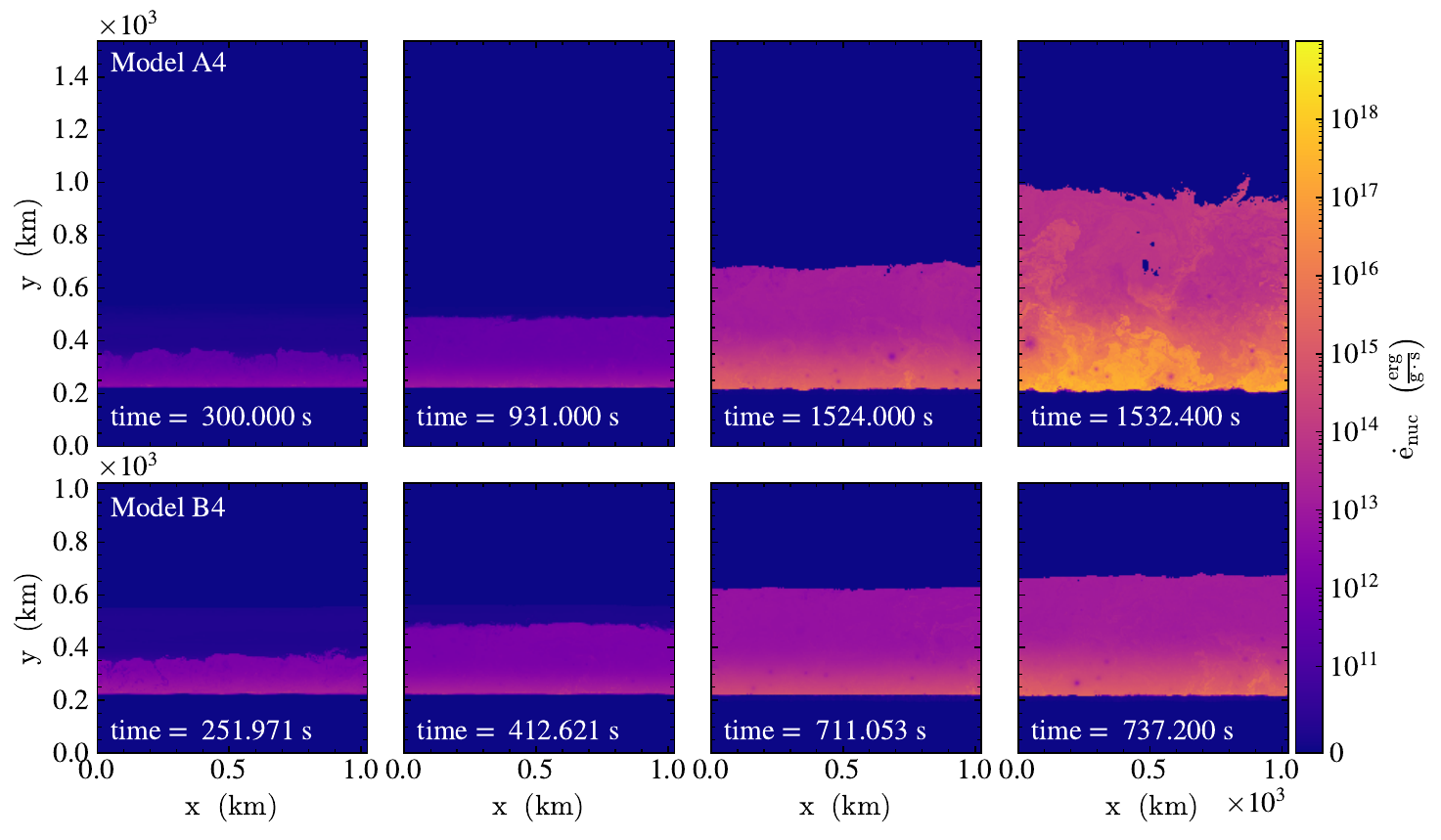}
\caption{A comparison of the evolution of model A4 and B4 at $0.4\,\mathrm{km}$, with two different size of the buffer-zone. (a) The first two rows, depict the evolution of the velocity magnitude of the A4 and B2 model respectively. (b) The third and fourth rows show the evolution of the nuclear specific energy rate of model A4 and B4 respectively.}
\label{fig:flow_compare} 
\end{figure*}

    \begin{table*}
        \centering
        \begin{tabular}{lllllrcl}  
            \hline \hline
            Model & $t_\gamma$ & $\Delta$ & $\Delta x_{\mathrm{max}}$ & $D_x \times D_y$ & \colhead{$N_x \times N_y$} & $T_{\mathrm{\mathrm{CEI}}}$ & Compute time\\ 
            \hline
            A4  & $1532.40\,\mathrm{s}$ &  $5\%$ & 0.4 km & $3072\times 1536\,\mathrm{km}$ & $1920\times 960$&$2.94\times 10^{8}\,\mathrm{K}$ & $3.734\times 10^{4}$ GPU-hr\\
            A8 & $1195.42\,\mathrm{s}$ &  $5\%$ & 0.8 km &  $3072\times 1536\,\mathrm{km}$ & $1920\times 960$ & $2.77\times 10^{8}\,\mathrm{K}$ & $1.175\times 10^{4}$ GPU-hr \\
            B2  &$704.19\,\mathrm{s}$ & $25\%$ &0.2 km & $2048\times 1024\,\mathrm{km}$ & $1280\times 640$&$1.92\times 10^8\,\mathrm{K}$ & $1.120\times 10^{5}$ GPU-hr \\
            B4 &$737.20\,\mathrm{s}$ & $25\%$ &0.4 km & $2048\times 1024\,\mathrm{km}$ & $640\times 320$&$1.87\times 10^8\,\mathrm{K}$ & $1.974\times 10^4$  GPU-hr \\
            B8  & $844.48\,\mathrm{s}$ &  $25\%$ & 0.8 km & $2048\times 1024\,\mathrm{km}$ & $1280\times 640$&$2.29\times 10^8\,\mathrm{K}$ & $1.899\times 10^{6}$ CPU-hr \\  
        \hline \hline  
        \end{tabular}  
        \caption{Summary of our simulations.  Here, $t_\gamma$ is the time just before matter crosses the top boundary (simulation ends), $\delta T$ is   initial temperature perturbation amplitude, $\Delta x_\mathrm{max}$ maximum resolution, $D_x\times D_y$ is the domain size, $N_x\times N_y$ is coarse grid number of cells, $T_{\mathrm{\mathrm{CEI}}}$ is the maximum value of the temperature just before the simulation end, and compute time is the total CPU / GPU hours used by the simulation.}
        \label{table:models}
    \end{table*}

\section{Results}

\subsection{Flow and nucleosynthesis evolution }

\begin{figure*}[t]
\epsscale{0.5}
\centering
\plotone{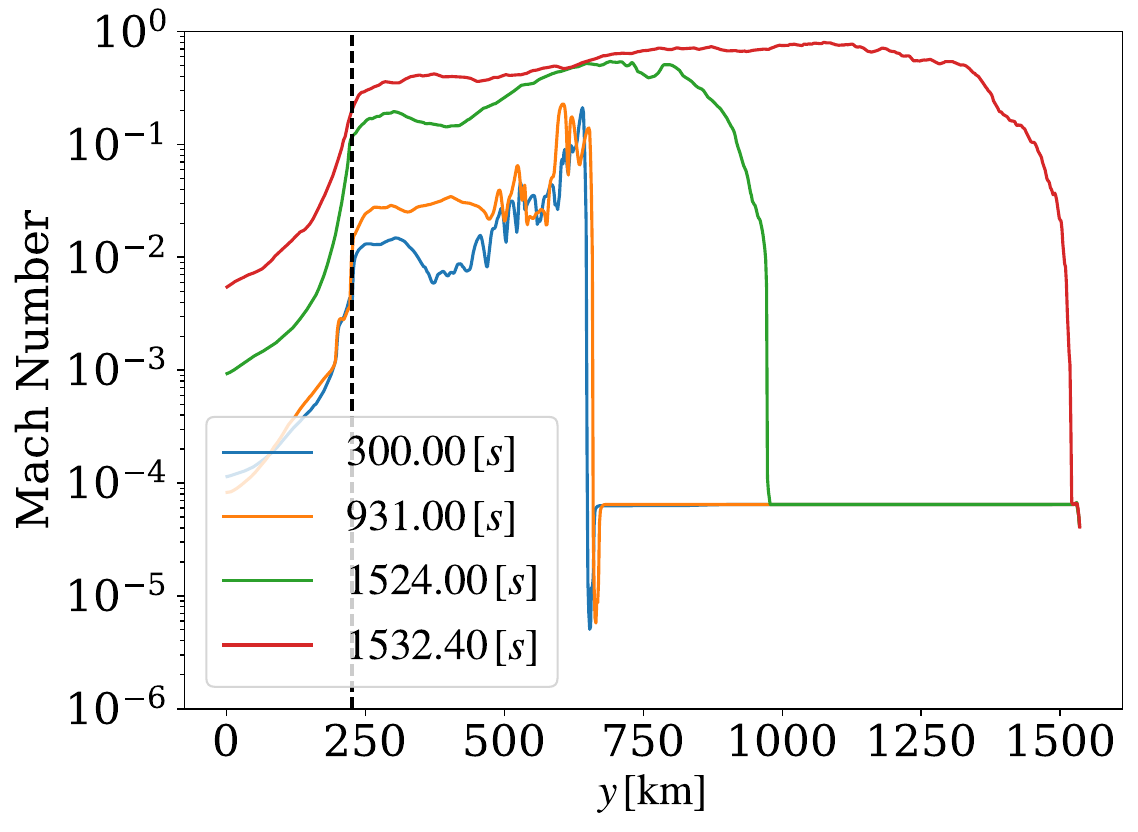}
    \caption{A comparison of the Mach number lateral average of model A4 at several times. The approximate Riemman solver of \texttt{CASTRO} is capable to handle cases where $\mathrm{Ma}\gtrsim 10^{-3}$. The
    vertical dotted line marks the location of the CEI.}
    \label{fig:mach}
\end{figure*}

Figure~\ref{fig:flow_compare} shows the evolution of the velocity field and the specific nuclear energy generation rate for models A4 and B4. As there are many similarities between the two models, we will focus on the evolution of model A4 and compare it against the remaining ones. Figure~\ref{fig:mach} shows a comparison of the lateral average of the Mach number for model A4 at different times. Note that for the convective envelope, the Mach Number ($\mathrm{Ma}$) increases from $\mathrm{Ma}\gtrsim 10^{-2}$, to $\mathrm{Ma}\sim 1$. Therefore, the magnitude of the convective velocity currents increases by two orders of magnitude by the time the simulation ends. 

The A4 simulation begins with a temperature perturbation of $5\%$, driving a violent burning phase that seeds strong velocity currents over a timescale of $\sim 20\mbox{--}50\,\mathrm{s}$. Once the temperature perturbation dissipates, the remaining velocity currents, with magnitudes of $|\mathbf{U}|\sim 10^6\,\mathrm{cm/s}$, move parallel to the CO layer, driving Kelvin-Helmholtz (KH) convective-unstable modes. These small vortex-like convective eddies have an initial size of $\Lambda\sim 200\,\mathrm{km}$, as seen at $t = 300 s$ in Figure~\ref{fig:flow_compare}. At this point, the timescale for the convective-turnover is about $\tau_{\mathrm{conv}}\sim 20\,\mathrm{s}$ and the specific nuclear energy generation rate is approximately $\dot{e}_\mathrm{nuc}\sim 10^{12}\mbox{--}10^{13} \,\mathrm{erg\,g^{-1}\,s^{-1}}$, suggesting that the energy evolution is primarily based on the hydrodynamic evolution of the density, pressure, and velocity fields, as convection becomes the more efficient energy transport mechanism at the CEI.

\begin{figure*}[t]
    \centering
    \plottwo{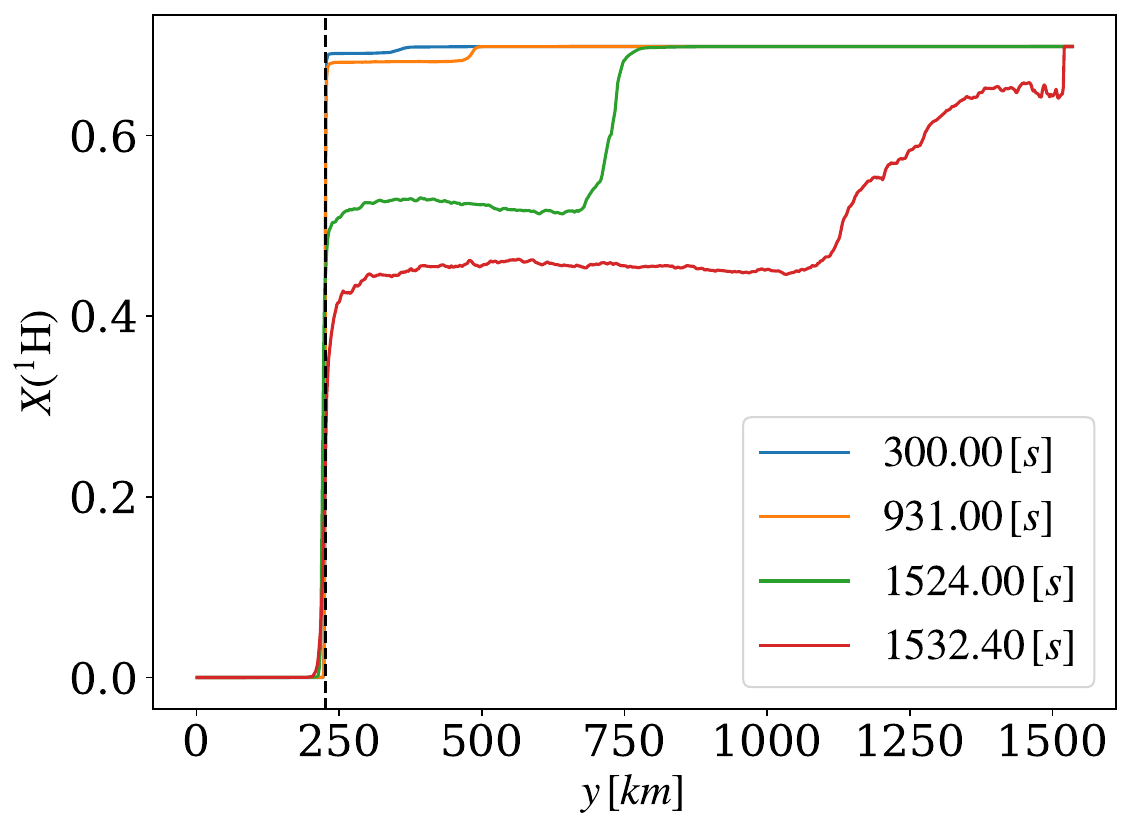}{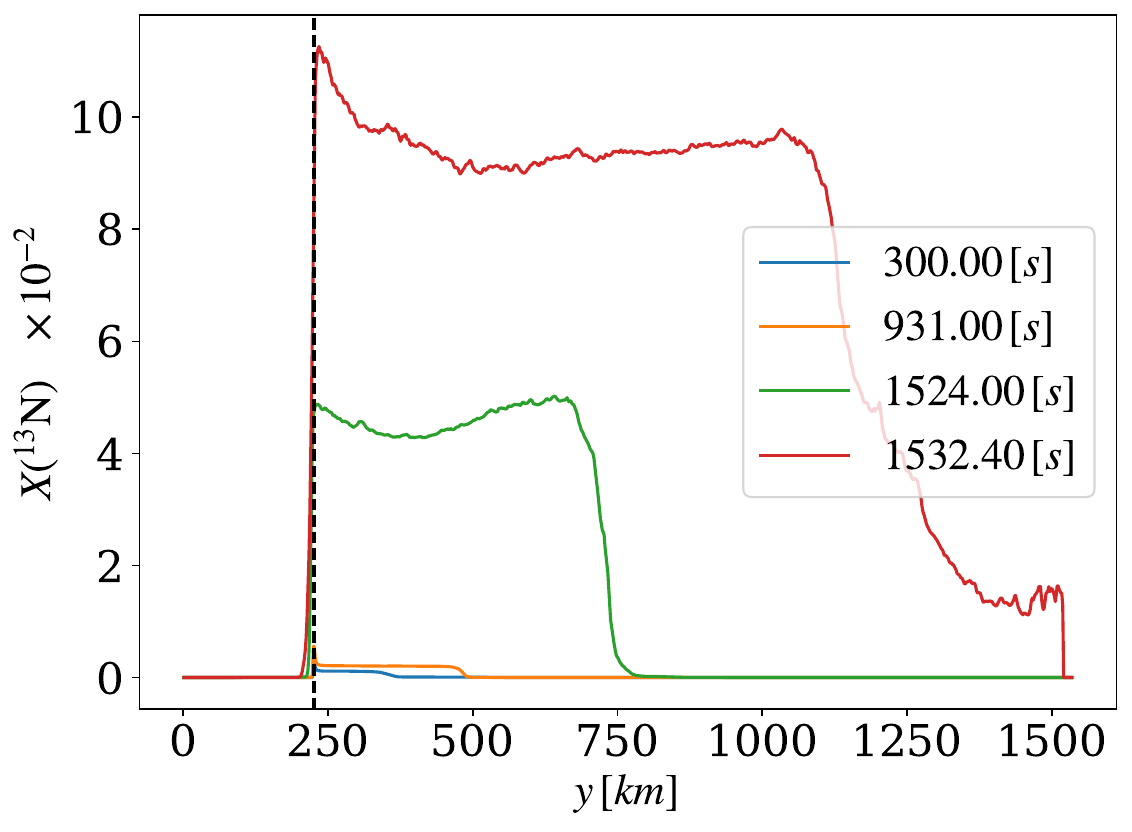}
    \caption{The evolution of the laterally-averaged mass fraction, weighted by mass, for the A4 model showing  (left) protons ($\isotm{H}{1}$) and (right) 
    $\isotm{N}{13}$. We see that as the proton mass fraction decreases near the CEI, a subtle valley forms in the $\isotm{N}{13}$ mass fraction profile toward  the end of the simulation.  The
    vertical dotted line marks the location of the CEI.}
    \label{fig:xn13}
\end{figure*}

\begin{figure*}[t]
    \centering
    \plottwo{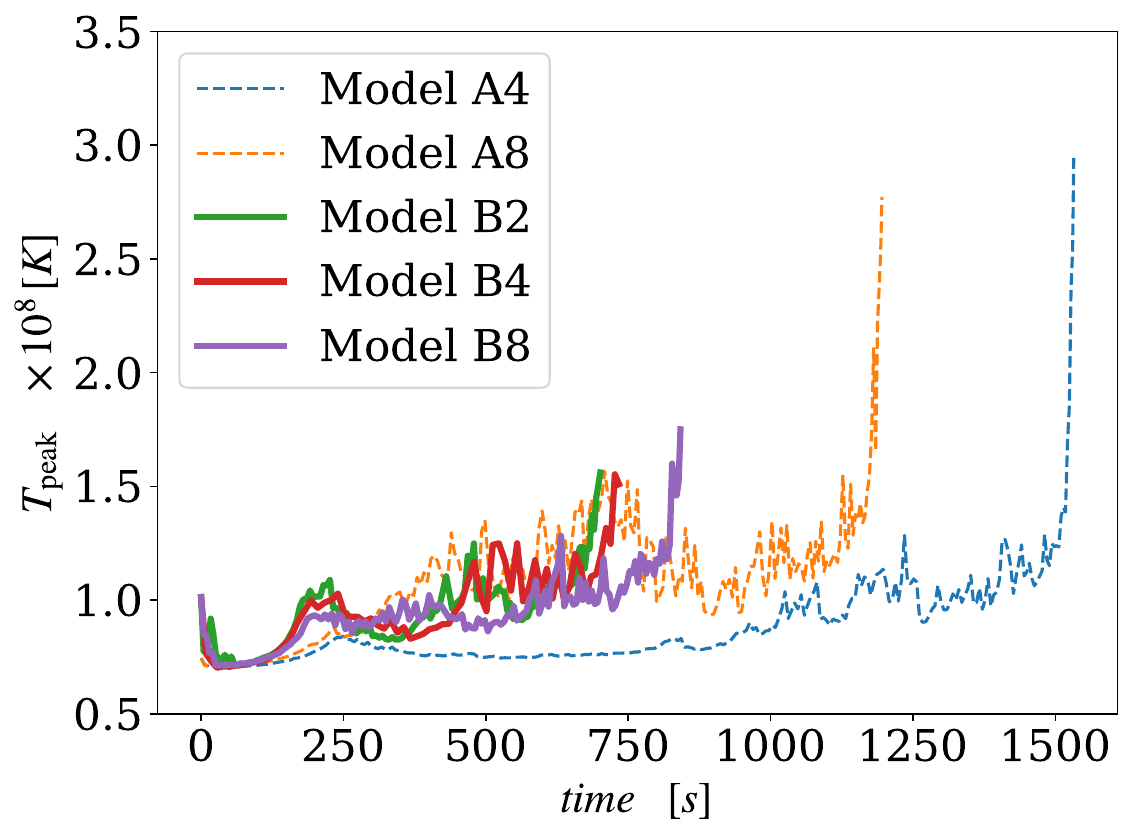}{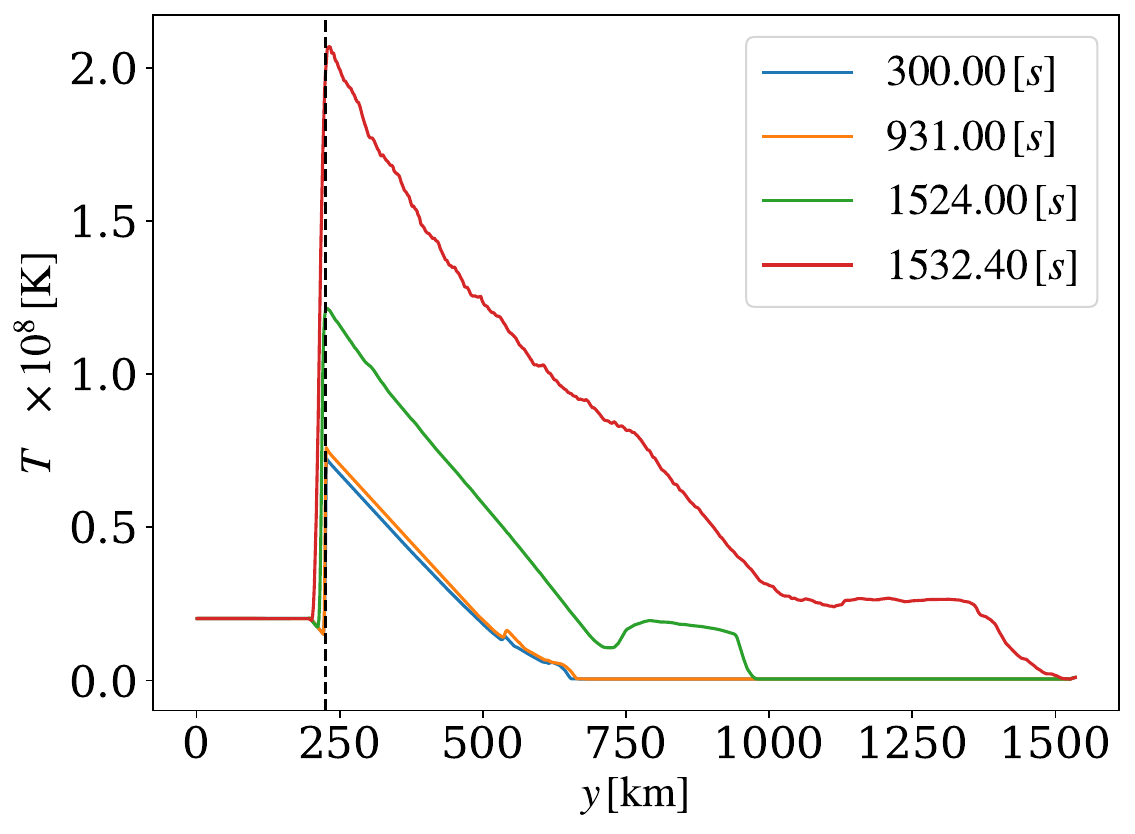}
    \caption{(left) The maximum temperature at the base of the accreted layer vs.\ time for all models; (right) The cell-volume-weighted lateral average of the temperature of model A4. The
    vertical dotted line marks the location of the CEI.}
    \label{fig:temp}
\end{figure*}

\begin{figure*}[t]
    \centering
    \plottwo{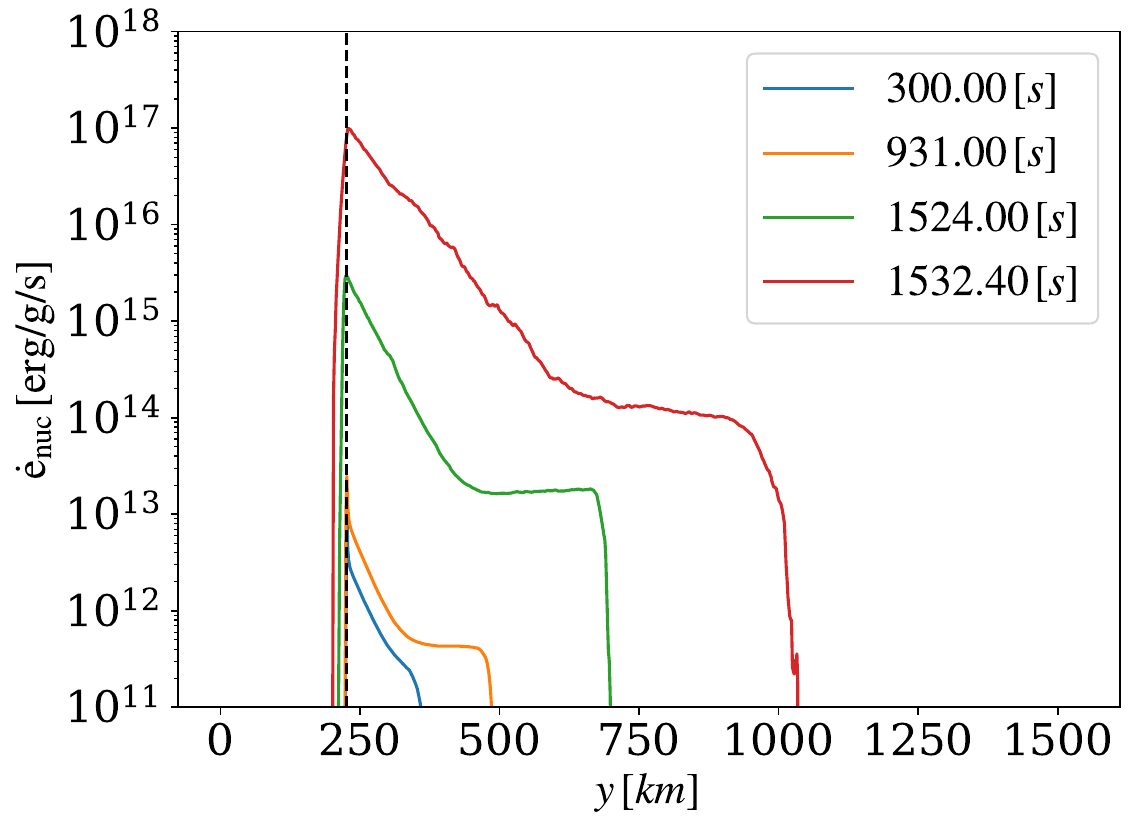}{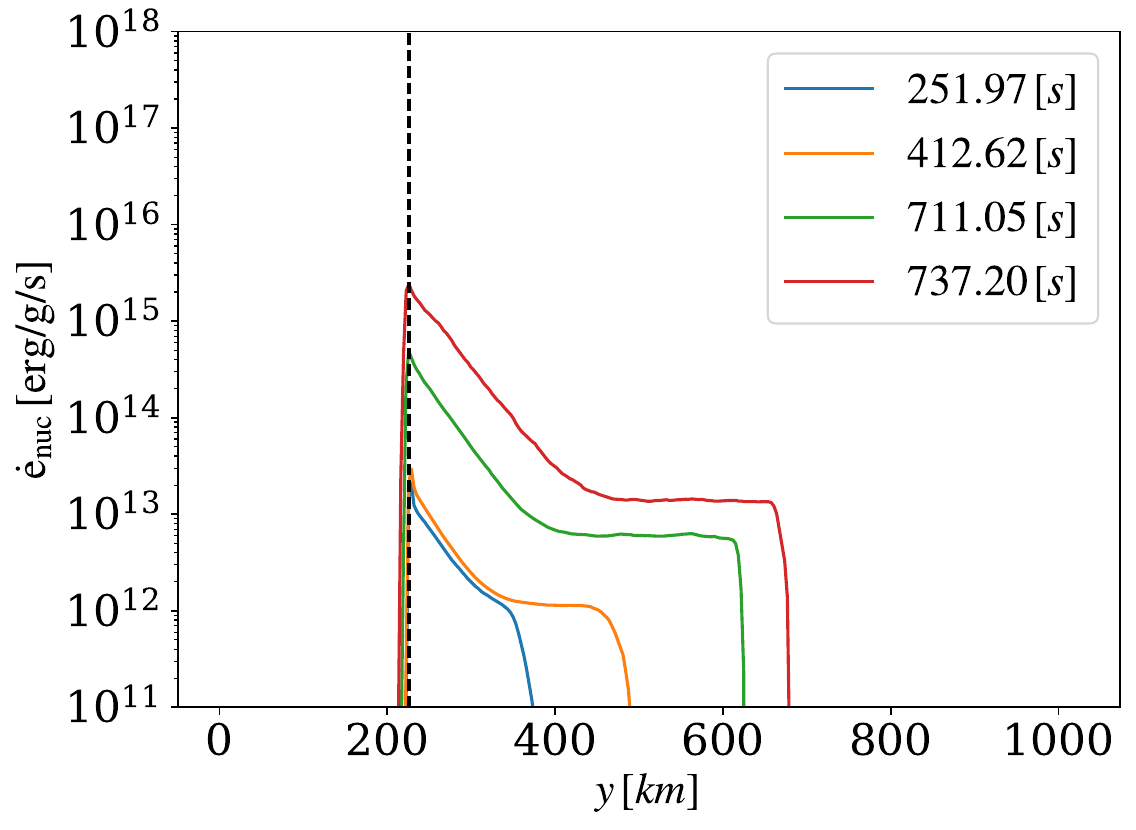}
    \caption{A comparison of the laterally-averaged profile of the specific nuclear energy generation rate between models A4 (left) and B4 (right).
    Both panels show the evolution up to $t_\gamma$ (which is significantly
    shorter for B4 than A4).  We observe a constant specific nuclear energy generation rate at the top of the envelope, dominated by the decay of $\beta^{+}$-unstable nuclei, like $\isotm{N}{13}$.  The
    vertical dotted line marks the location of the CEI.}
    \label{fig:enuc}
\end{figure*}

As the simulation evolves ($t=931\,\mathrm{s}$ in Figure~\ref{fig:flow_compare}), the convective eddies merge, creating larger eddies ($\Lambda\sim 300\,\mathrm{km} $) with higher velocity ($|\mathbf{U}|\sim 10^7\,\mathrm{cm/s}$), reducing the convective-turnover timescale to $\tau_\mathrm{conv} \sim 3\,\mathrm{s}$. At
this point, fresh $\isotm{C}{12}$ and $\isotm{O}{16}$ spreads
from the CO layer to the accreted layer, creating suitable
conditions $p$-captures, and $\beta^{+}$-decays
near the CEI through the CNO-cycle. Following the arguments of \cite{wiescher:1999}, and using \texttt{pynucastro} to compute the lifetime of all the CNO-nuclei in our reaction network against $p$-captures and $\beta^{+}$-decays at $\rho \sim 10^3\,\mathrm{g/cm^3}$, $T=7.22\times 10^7\,\mathrm{K}$, and proton mass fraction $X_p\sim 0.70$, we find:
\begin{eqnarray}
     \tau^{p} &=& \dfrac{1}{\rho X_p N_A\langle \sigma v \rangle_{\mathrm{CNO}(p,\gamma)}}\sim 10^{3}\mbox{--}10^{5}\,\mathrm{s}, \label{eq:p_rate} \\
     \tau^{\beta^{+}} &=& \dfrac{\log(2)}{\lambda_{\beta^{+}}}\sim 10\mbox{--}100\,\mathrm{s}, \label{eq:beta_rate}
\end{eqnarray}
where $N_A\langle \sigma v\rangle_{\mathrm{CNO}(p,\gamma)}$ is the magnitude of each $\mathrm{CNO}$-nuclei $p$-capture reaction rate, and $\lambda_{\beta^{+}}$ is the magnitude of each $\beta^{+}$-decay rate of the reaction network. Looking
at all the $p$-captures, the slowest reactions are $\isotm{C}{12}(p,\gamma)\isotm{N}{13}$ in the CN-cycle and $\isotm{O}{16}(p,\gamma)\isotm{F}{17}$ in the ON-cycle. These reactions are the first $p$-capture onto the $\isotm{C}{12}$ and $\isotm{O}{16}$ nuclei ingested into the accreted layer. Because $\tau^{\beta^{+}}\ll \tau^{p}$, the main energy contribution to the accreted layer comes from the decay of the existing $\beta^{+}$-unstable nuclei that are uniformly mixed across the entire accreted layer, since
$\tau_\mathrm{conv}\lesssim\tau^{\beta^{+}}$.  This increases the presence of
these $\beta^{+}$-unstable nuclei ($\isotm{N}{13}$, $\isotm{O}{14}$, $\isotm{O}{15}$, and $\isotm{F}{17}$) near the top of
the accreted layer (Figure~\ref{fig:xn13}) as the temperature of the CEI increases. 
As these nuclei move farther from the CEI, they encounter lower temperatures ($\lesssim 10^{7}\,\mathrm{K}$, see, Figures \ref{fig:inital_models} and \ref{fig:temp}), where $\beta^{+}$-decay reactions are the only option. Therefore, close to
the top of the accreted layer, the nuclear specific energy rate becomes
independent of temperature, depending only on the
metallicity of carbon, nitrogen and oxygen, $Z_\mathrm{CNO}$, as discussed in \cite{glasner:1997}. 
In Figure~\ref{fig:enuc}, we see that the energy generation rate as a function of height flattens as we get close to the top of the accreted layer, and that its level increases in time as as $Z_\mathrm{CNO}$ increases
through mixing. 

As the $\beta^{+}$-unstable nuclei
decay, the energy released is distributed uniformly throughout entire
accreted layer, modifying the pressure and density, and slightly
increasing the average temperature of the accreted layer (Figure \ref{fig:temp}). The nuclei
$\isotm{C}{13}$ and $\isotm{O}{17}$, produced via $\isotm{N}{13}(\beta^+\nu)\isotm{C}{13}$ and
$\isotm{F}{17}(\beta^+\nu)\isotm{O}{17}$, are stable against $\beta^{+}$-decays, meaning  they can only participate in $p$-captures once they return to CEI through the
convective-eddy currents, and the process repeats. These
particular modifications in the pressure and density fields produce
additional currents that initiate an inverse turbulent cascade.  We note that
the kinetic  energy cascade in three dimensions moves from large to small eddies, while conservation of vorticity in two dimensions reverses its direction \citep{ouellette:2012}. Therefore, the inverse turbulent cascade is a consequence of the dimensionality of our simulations, implying that a correct description of the cascading process requires a 3D simulations.

At $t=1524.00\,\mathrm{s}$, the size of the convective-eddies are
comparable to the size of the accreted layer (Figure~\ref{fig:flow_compare}), with $\Lambda\sim
400\,\mathrm{km}$, and the magnitude of the velocity currents increases to
$|\mathbf{U}|\sim 10^8\,\mathrm{cm/s}$, further reducing of the convective-turnover time to $\tau_c\sim 0.2\,\mathrm{s}$.  At this point, from Figure \ref{fig:enuc},
we also see a substantial increase in the specific nuclear energy generation rate, up to $\dot{e}_\mathrm{nuc} \sim 10^{15}\,\mathrm{erg}\,\mathrm{g^{-1}}\,\mathrm{s^{-1}}$. The continued energy released from the many $p$-captures and $\beta^{+}$-decays, pushes the temperature of the CEI above the threshold $\sim 10^8\,\mathrm{K}$, transitioning from the CNO to the hot-CNO cycle, making $p$-captures and $\beta^{+}$-decay equally likely. Again using \texttt{pynucastro} to evaluate the $p$-captures and $\beta^{+}$-decays (Eqs. \ref{eq:p_rate} and \ref{eq:beta_rate}), at thermodyamic conditions
representative of this stage, $\rho \sim 10^{3}\,\mathrm{g/cm^3}$, $X_p\sim 0.50$ (see Figure \ref{fig:xn13}), and $T=1.21\times 10^8\,\mathrm{K}$ (see Figure \ref{fig:temp}) we find:
\begin{eqnarray}
     \tau^{p} &\sim& 10\mbox{--}1000\,\mathrm{s}, \label{eq:p_rate_m}\\
     \tau^{\beta^{+}} &\sim& 10\mbox{--}100\,\mathrm{s}. \label{eq:beta_rate_m}
\end{eqnarray}
showing that now $\tau_\mathrm{conv}\ll \tau^{p} \sim \tau^{\beta^{+}}$. Therefore, nuclei such as $\isotm{N}{13}$ may produce $\isotm{C}{13}$ and $\isotm{O}{14}$ through the reactions $\isotm{N}{13}(\beta^{+}\nu)\isotm{C}{13}$ and $\isotm{N}{13}(p,\gamma)\isotm{O}{14}$, and uniformly mix all the reactants and products of these reactions through the entire accreted layer. 
Figure~\ref{fig:xn13} shows the proton ($\isotm{H}{1}$) and $\isotm{N}{13}$ mass fractions near the CEI, indicating a lack of protons and $\isotm{N}{13}$ near the CEI, as more $\isotm{N}{13}$ engage in $p$-captures.

Finally, at $t=1532.40$, just before the top of the atmosphere reaches the upper boundary of our computational domain, $p$-captures become substantially more likely than $\beta^{+}$-decays. Reevaluating the proton capture and $\beta^+$-decay timescales at these thermodynamic conditions, $\rho=10^{3} \,\mathrm{g/cm^3}$, $X_p\sim 0.45$, and $T=2.07\times 10^8\,\mathrm{K}$, we now see:
\begin{eqnarray}
     \tau^{p} &\sim& 0.01\,\mbox{--}\,1\,\mathrm{s},\\
     \tau^{\beta^{+}} &\sim& 10\,\mbox{--}\,100\,\mathrm{s}. 
\end{eqnarray}
showing that $\tau^p \ll \tau^{\beta^{+}}$. The slowest $\beta^{+}$ decays are the reactions $\isotm{N}{13}(\beta^{+}\nu)\isotm{C}{13}$ and $\isotm{O}{15}(\beta^{+}\nu)\isotm{N}{15}$ located in the CN- and ON-branches respectively. Therefore, between these two waiting points, a rapid surge of energy produced by $p$-captures is released to the accreted layer, producing the necessary conditions for the explosive expansion of the atmosphere.

\begin{figure*}[t]
    \centering
    \plottwo{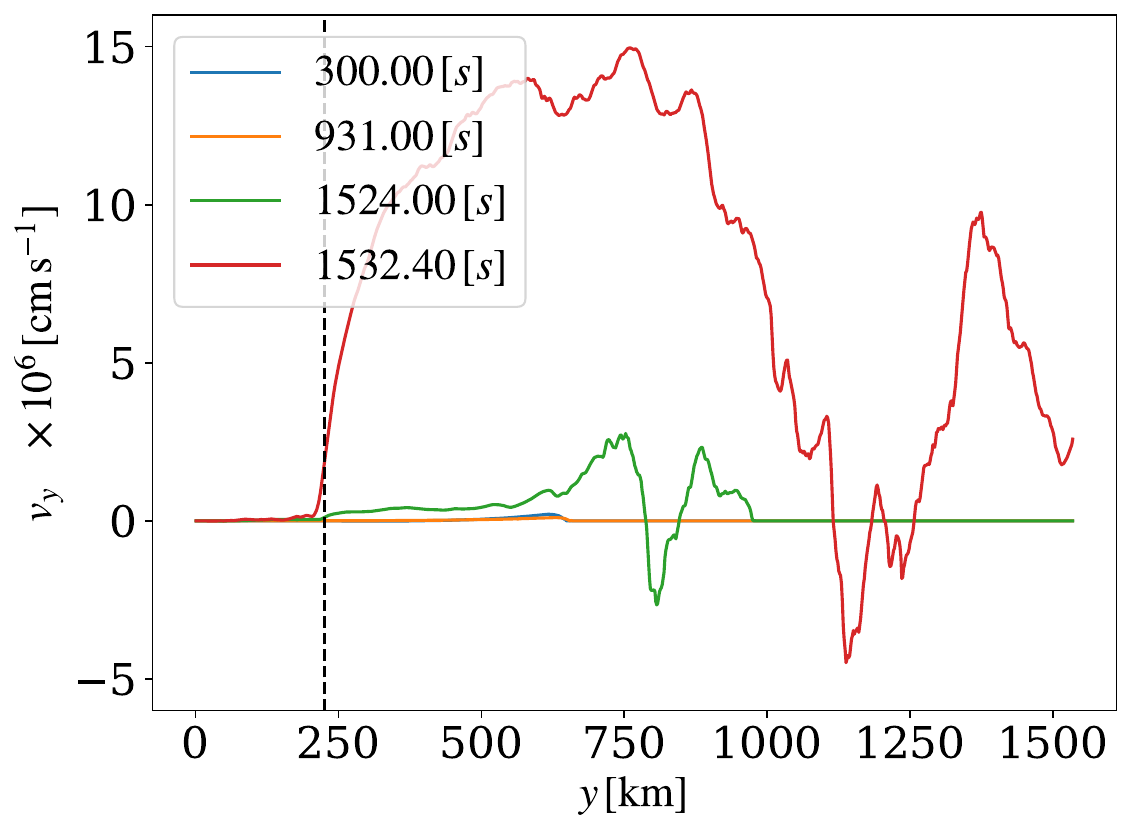}{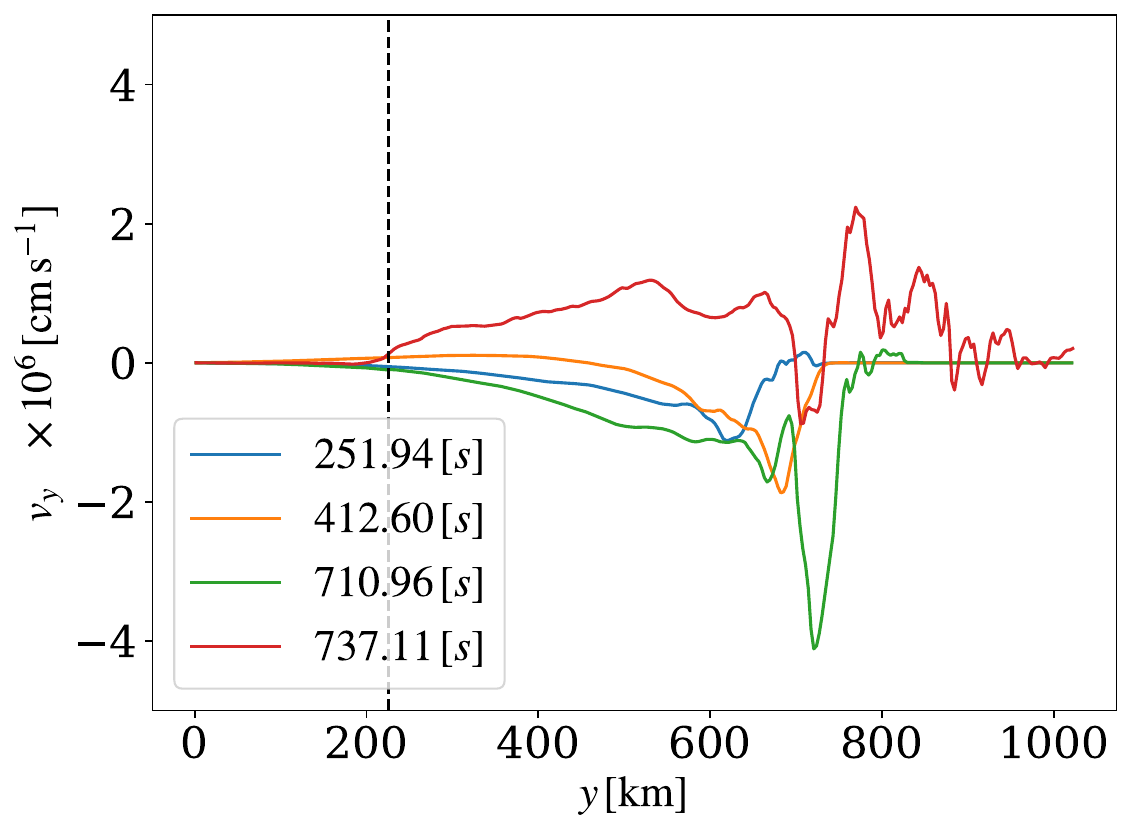}
    \caption{A comparison of the laterally-averaged profile of the radial (vertical) velocity for models A4 (left) and B4 (right). The main difference between the two models is the location of the peak. In model A4, the peak is reached at the CEI, suggesting that the energy provided by the double $p$- captures powers the explosive expansion of the envelope. However in model B4, the peak is directed downwards and located at the interface between the accreted layer and the buffer-zone, suggesting the presence of convective-eddies that are still in place.  The
    vertical dotted line marks the location of the CEI.}
    \label{fig:velocity}
\end{figure*}

\begin{figure*}[t]
    \centering
    \plottwo{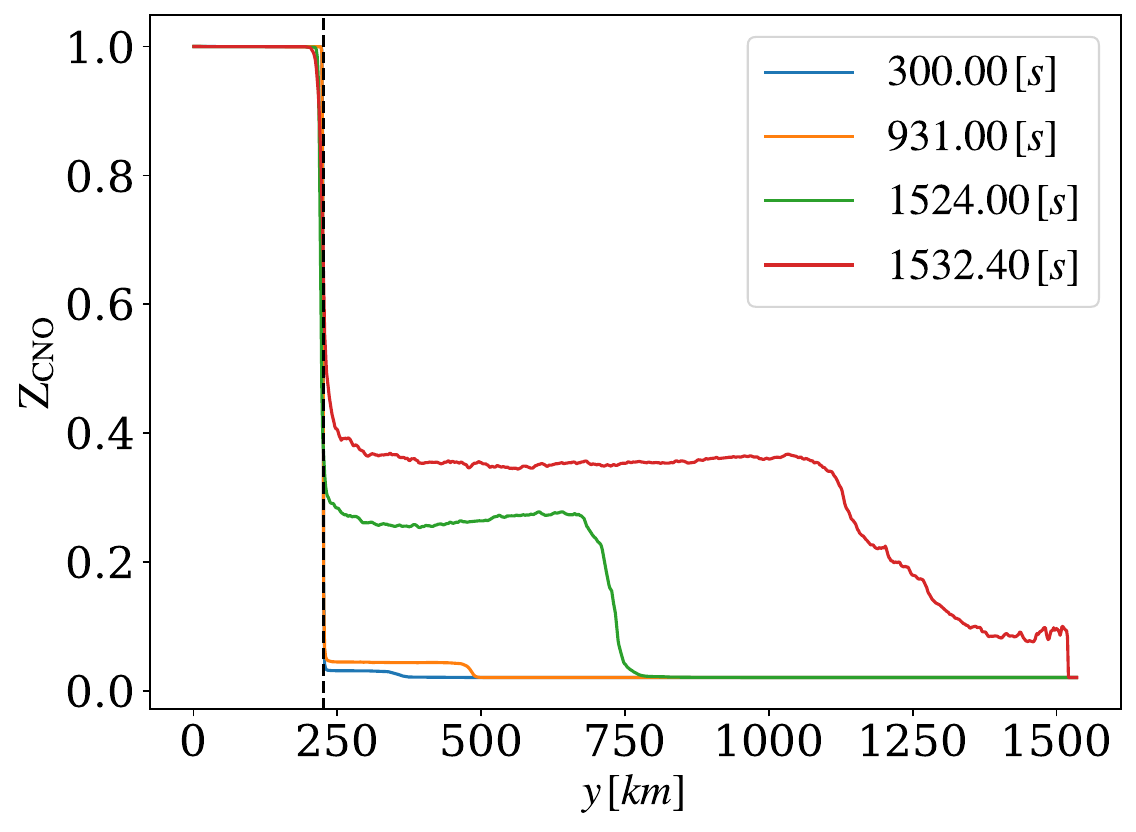}{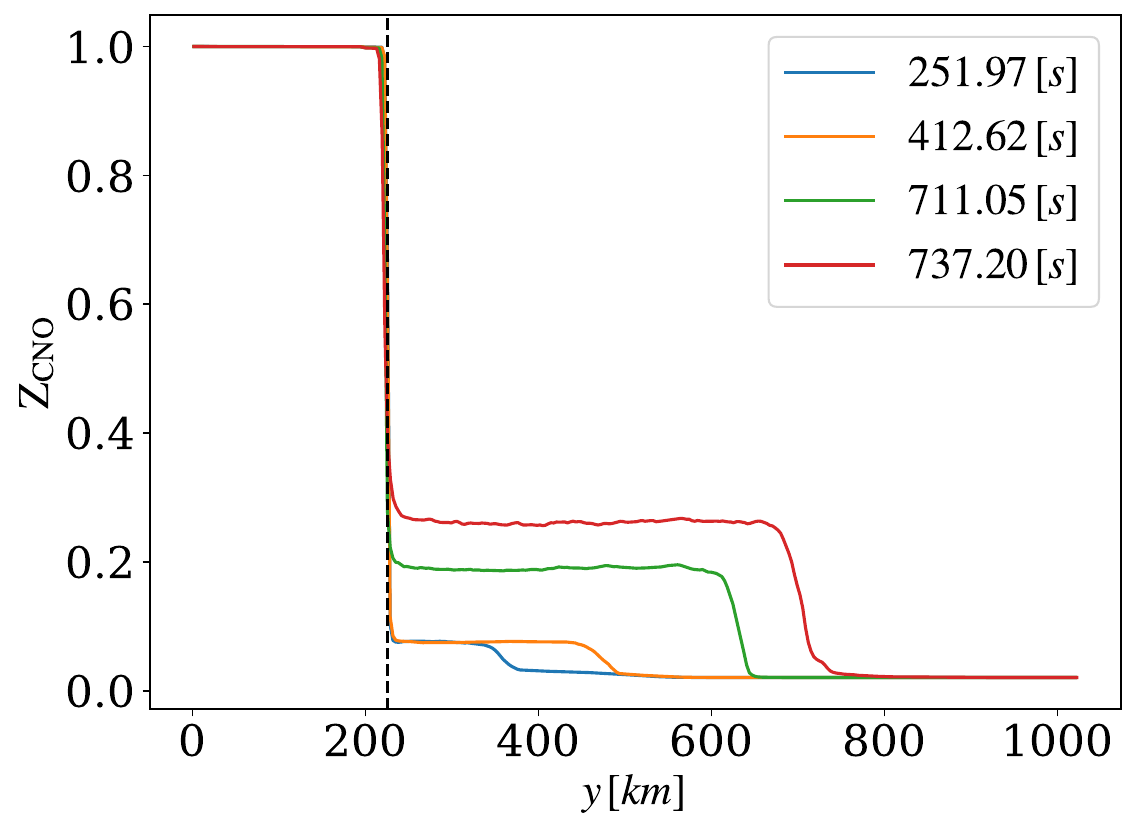}
    \caption{A depiction of the lateral average evolution of the sum of the $\isotm{C}{12,13}$, $\isotm{N}{13,14,15}$ and $\isotm{O}{13,14,15,16}$ nuclei mass fractions $Z_\mathrm{CNO}$ for models A4 (left) and B4 (right). The vertical dotted line marks the location of the CEI.}
    \label{fig:xcno}
\end{figure*}

\begin{figure*}[t]
    \centering
    \plottwo{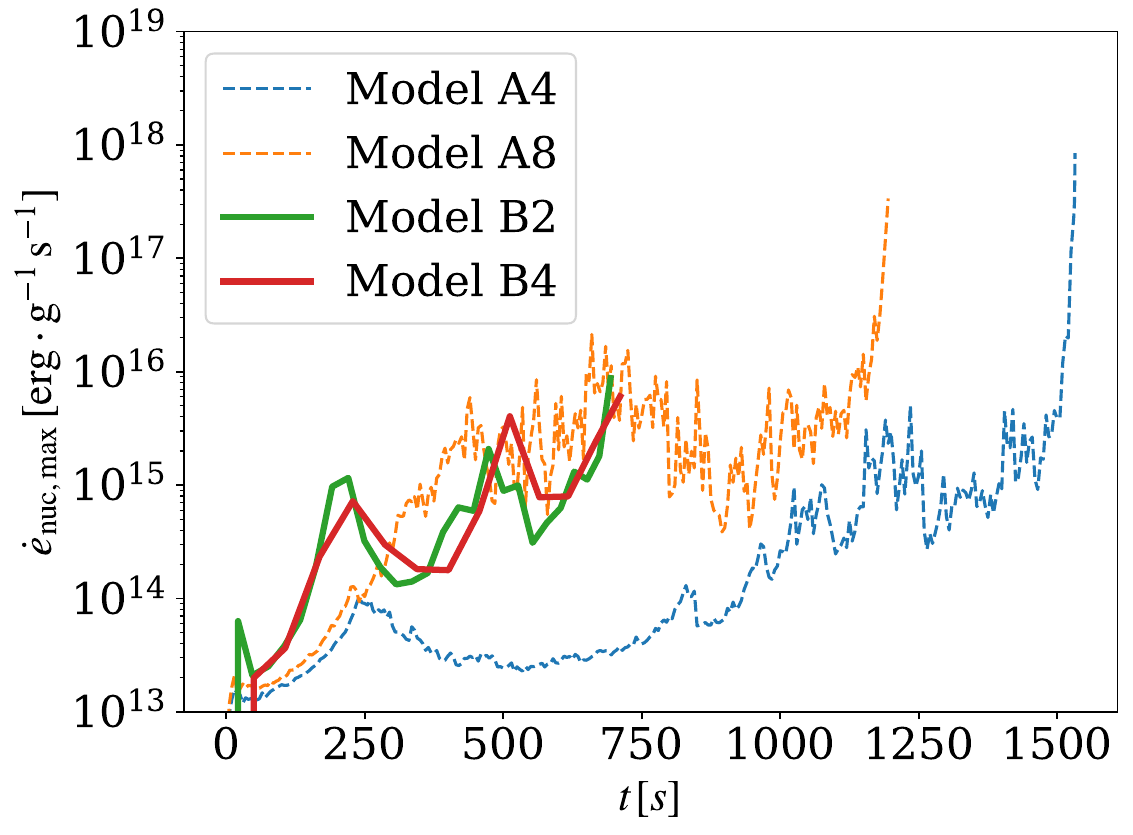}{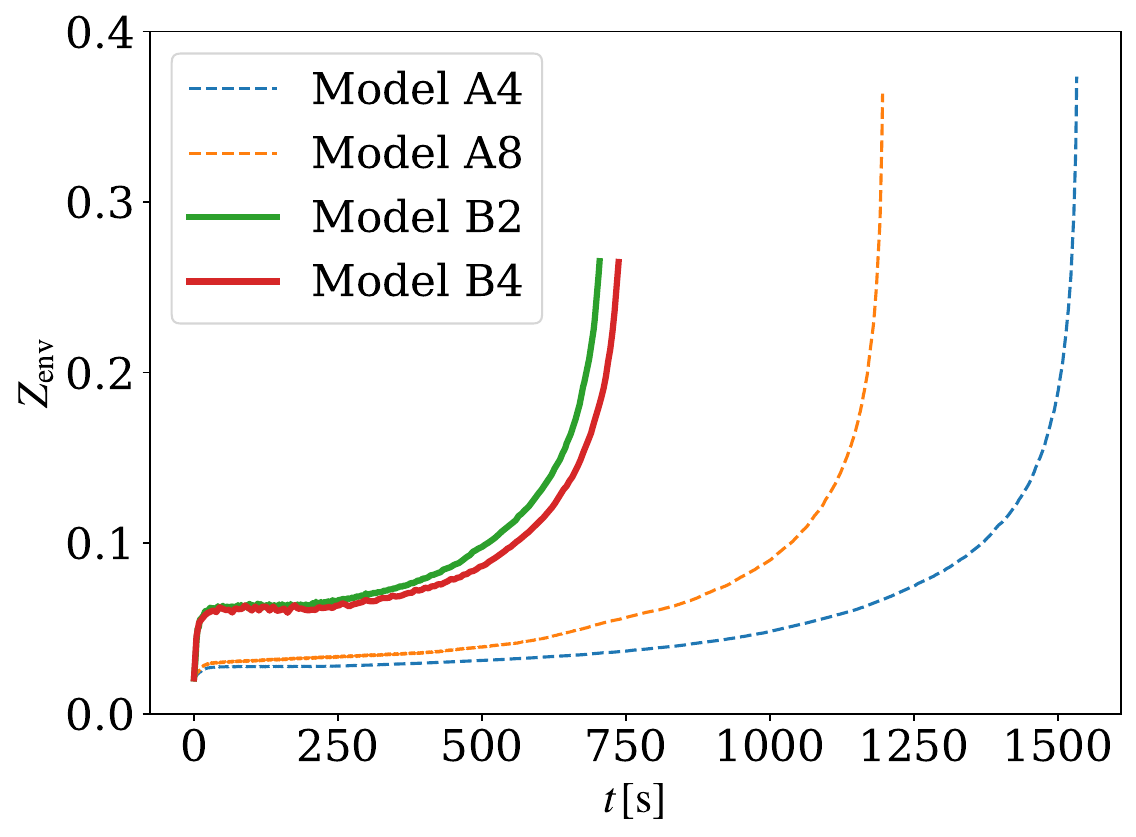}
    \caption{A comparison of the time evolution of the maximum specific nuclear energy generation rate, $\dot{e}_{\mathrm{nuc, max}}$ (left) and the metallicity content of the envelope (right).}
    \label{fig:enuc_t}
\end{figure*}

The evolution of the radial velocity profile of models A4 and B4 (Figure~\ref{fig:velocity}), show a significant difference between them as they reach $t_\gamma$. From model A4, a broad peak is observed centered at the CEI with a magnitude of $v_{y,\mathrm{peak}} \sim 1.5 \times 10^{7} \,\mathrm{cm/s}$ directed outward, while in model B4, we see a sharp peak at the interface between the buffer-zone and the accreted layer with a magnitude of $v_{y,\mathrm{peak}}\sim 4.0\times 10^{6}\,\mathrm{cm}$ directed downwards. This pattern in the model B4 radial velocity profile suggests that convective-eddies are still pushing CNO nuclei upwards in the accreted layer without giving enough time for the last convective-turnovers to let the $p$-captures and $\beta^{+}$-decays to occur and increase the temperature of the CEI. Therefore, we are not able to see from B4 (and consequently from models B2 and B8) the last stage of the flow evolution of model A4 (and A8) at $t=1532.40\,\mathrm{s}$ in Figure~\ref{fig:flow_compare}. This
comparison stresses the importance of having a large enough buffer-zone to allow the accreted layer to expand.

The metallicity enrichment in models A4 and B4 is shown in Figure~\ref{fig:xcno}. The enrichment of CNO metalicity in model A4 ($Z_\mathrm{CNO}\sim 0.3\mbox{--}0.4$) is much higher than in B4 ($Z_\mathrm{CNO}\sim 0.2\mbox{--}0.3$), even though model B4 used a larger initial temperature perturbation.  Again, this highlights the importance of an extended buffer-zone that allow a significant space for combustion to occur and dilute the influence of the boundary conditions as the simulation evolves. 

\subsection{Temperature and energy evolution}

Figure~\ref{fig:temp} shows the maximum temperature at the CEI vs.\ time of all our runs as well as the lateral average of the temperature (weighted by cell volume) for model A4. From model A4, after the appearance of the first KH instabilities, the temperature drops to its initial value, remaining almost constant for approximately $500\,\mathrm{s}$.  This model reaches the non-linear temperature increase
associated with TNR stage much later than the B-series models or even model A8. This nearly-constant temperature period allows model A4 to realize a smooth transition between the CNO cycle to the hot-CNO cycle at the CEI, dissipating all excess heat generated by the decay of the $\beta^{+}$-unstable nuclei to the top of the accreted layer. Therefore, the initial perturbation creates the necessary conditions for the initial $p$-captures to increase the mass fraction of the $\beta^{+}$-unstable nuclei that the CNO cycle requires to operate. The comparison of models B2 and B4 show that the temperature increase appears converged, justifying the resolution we chose for our main simulation, $\Delta x_\mathrm{max} =0.4\,\mathrm{km}$.  This convergence is even despite the enhanced initial perturbation size of $25\%$, compared with the $5\%$ of models A4.

A comparison of the specific nuclear energy generation rate evolution among all the models is shown in Figure~\ref{fig:enuc_t}, with a rapid and enhanced growth in models B2 and B4, from  $\lesssim 10^{11}$ to $\sim 10^{13}\mbox{--}10^{14}\,\mathrm{erg\,g^{-1}\,s^{-1}}$. Although this initial surge is artificially generated by the enhanced $25\%$ temperature perturbation-size that acts only at the initial timestep, there are important consequences for the evolution of the decay of $\beta^{+}$ unstable nuclei. From the sensitivity studies in \cite{casanova:2011a}, as long as the perturbation is applied only at the initial step, its size and shape does not significantly influence the evolution flow. From our results, we confirm that although the flow remains unchanged, this particular initial peak in the energy production creates the conditions for a significant metallicity enhancement after the initial perturbations completely dissipates. 
Figure~\ref{fig:enuc_t} also shows a surge of the metallicity from the initial value of $Z=0.02$ to $Z\gtrsim 0.05$, which constitutes a substantial increase after the perturbations dissipate completely. Therefore, the effects of an increased perturbation only modify the initial metallicity, which may produce substantial effects on the energetic contribution near the top of the accreted layer.  This is because the $\beta^{+}$-decays are the only set of reactions that operate at lower temperatures, depending only on the composition \citep{glasner:2007}, reducing the required time for the eddy-convective currents to transport the energy out of the CEI. 

\section{Discussion and Conclusions}

In this paper we focused on the study of the sensitivity of two-dimensional
simulations of novae, starting with the T7 model of \citet{glasner:1997}.
We demonstrated the advantages of a buffer-zone between the top of the accreted layer and the upper boundary of the domain and the maximum resolution employed in our models. 

Our model A4 reproduces the beginning of the TNR, consistent with the observed values of temperatures close to $T\sim 2.1\times 10^8\,\mathrm{K}$ and enhanced values of the metallicity $Z\sim 0.30\mbox{--}0.40$, slightly higher than the $Z\sim 0.20\mbox{--}0.30$ values from \cite{casanova:2010,casanova:2011a,casanova:2011b,casanova:2016,casanova:2018}, and our models B2, B4, and B8. The convergent trend of models B2 and B4 in both temperature and specific nuclear energy generation rate validates our choice for the maximum resolution $\Delta x_\mathrm{max}=0.4\,\mathrm{km}$, used in model A4. Furthermore, by comparing the flow evolution, specific nuclear energy generation rate, temperature, vertical velocity, and metallicity profiles between models A4 and B4, we conclude that the presence of a buffer-zone on the top of the accreted layer is essential to capture the thermodynamic state evolution at the CEI leading to the TNR. The flow and nucleosynthesis evolution of model A4 shows a tight correlation between the timescale of the two nuclear dominant processes: the $\beta^{+}$-decays and $p$-captures, and the timescale of convection as the simulation evolves. We observe that the convective time scale, $\tau_\mathrm{conv}$, is significantly reduced from an initial $\tau_\mathrm{conv}\sim 20\,\mathrm{s}$ where $\tau^\beta\lesssim \tau^p$ to $\tau_\mathrm{conv}\sim 0.2\,\mathrm{s}$ where $\tau^\beta\gg \tau^p$. This transition on which nuclear process dominates in the different stages towards the TNR suggests an important relationship between the $\beta^{+}$-decays and $p$-captures, and the evolution of the convective eddies.

While our reaction network includes the chain of rates $\isotm{C}{12}(p,\gamma)\isotm{N}{13}(p,\gamma)\isotm{O}{14}(\beta^{+}\nu)\isotm{N}{14}$, the analogous chain of rates $\isotm{O}{16}(p,\gamma)\isotm{F}{17}(p,\gamma)\isotm{Ne}{18}(\beta^{+}\nu)\isotm{F}{18}$ is not present. Therefore, two protons are removed from the burning of $\isotm{C}{12}$, while only one proton is removed by $\isotm{O}{16}$. This asymmetry, between the $\isotm{C}{12}$ and $\isotm{O}{16}$ nuclei, may suggest a deficit in the overall metallicity. In addition, the relative abundance between the CNO-nuclei, $\isotm{F}{17}$, and $\isotm{Ne}{18}$ may provide an estimate on the size of the CNO breakout for this problem \citep{koon:2003}.  This can be addressed with a larger network in future simulations.

The Mach number for these simulations
was reasonable for a fully compressible code.  However, if we were to start earlier in the evolution, we might explore low Mach methods (like \citealt{maestroex}), or consider
Riemann solvers applicable to lower Mach numbers \citep{hlld}.

Because our initial model is an snapshot of a one-dimensional profile that uses MLT to reproduce the convective energy transport, there are non-zero velocities initially.  When we map this into two-dimensions, we cannot preserve the velocity field, resulting in the initial transient seen in our simulations.  We can lessen this effect in the future by adopting the implementation of a self-consistent initial convective velocity field, described in \citet{castro-massive-stars}. 

Finally, by following the same guidelines for the construction of the presented ``A-series" models, we ran 100 steps of two 3D models, with a domain of $1000\,\mathrm{km}\times 1000\,\mathrm{km}\times 1500\,\mathrm{km}$, and resolutions of $0.8\,\mathrm{km}$ and $0.4\,\mathrm{km}$. For the $0.8\,\mathrm{km}$ resolution, we estimated a computational cost of $\sim 10^5\,\mathrm{GPU\mbox{-}hr}$, while a resolution of $0.4$ may require $\sim 10^6\,\mathrm{GPU\mbox{-}hr}$, using $256$ nodes on the OLCF Frontier machine. Therefore, through a compromise in the size of the domain and the resolution employed by these runs, feasible 3D runs can be performed. 

\section*{Acknowledgments}

We thank Ami Glasner for providing us with the 1D initial models
and for helpful comments on this paper. Also we thank Eric Johnson for his insightful remarks and help with the code. Castro and the AMReX-Astrophysics suite are freely available at
 \url{https://github.com/AMReX-Astro/}.  All of the code and problem
 setups used here are available in the git repo. All of the metadata and global diagnostics are available in a zenodo record as \cite{smith:2025}. The work at Stony Brook was supported by DOE/Office of Nuclear Physics grant DE-FG02-87ER40317, and an award from the Chilean Government, Minisitry of Science and Technology, through the ANID grant ``Beca de Doctorado Igualdad de Oportunidades Fulbright (BIO)", number 56160017, and sponsored by the U.S Department of State, Fulbright, grant number PS00280789. This research used resources of the National Energy Research Scientific Computing Center (NERSC), a Department of Energy Office of Science User Facility using
 NERSC award NP-ERCAP0027167. An award for computer time was provided by the U.S. Department of Energy’s (DOE) Innovative and Novel Computational Impact on Theory and Experiment (INCITE) Program.  This research used resources from the Argonne Leadership Computing Facility, a U.S. DOE Office of Science user facility at Argonne National Laboratory, which is supported by the Office of Science of the U.S. DOE under Contract No. DE-AC02-06CH11357.

\bibliographystyle{yahapj}
\bibliography{nova}

\begin{thebibliography}{}
\providecommand\natexlab[1]{#1}
\providecommand\JournalTitle[1]{#1}

\bibitem[{Almgren {et~al.}(2020)Almgren, Sazo, Bell, Harpole, Katz, Sexton, Willcox, Zhang, \& Zingale}]{castro_joss}
Almgren, A., Sazo, M.~B., Bell, J., {et~al.} 2020, \href{http://dx.doi.org/10.21105/joss.02513}{\JournalTitle{Journal of Open Source Software}, 5, 2513}

\bibitem[{Almgren {et~al.}(2010)Almgren, Beckner, Bell, Day, Howell, Joggerst, Lijewski, Nonaka, Singer, \& Zingale}]{castro}
Almgren, A.~S., Beckner, V.~E., Bell, J.~B., {et~al.} 2010, \href{http://dx.doi.org/10.1088/0004-637x/715/2/1221}{\JournalTitle{ApJ}, 715, 1221}

\bibitem[{Bell {et~al.}(1989)Bell, Colella, \& Trangenstein}]{bell:1989}
Bell, J.~B., Colella, P., \& Trangenstein, J.~A. 1989, \href{http://dx.doi.org/https://doi.org/10.1016/0021-9991(89)90054-5}{\JournalTitle{Journal of Computational Physics}, 82, 362}

\bibitem[{Brandenburg {et~al.}(2009)Brandenburg, Svedin, \& Vasil}]{brandenburg:2009}
Brandenburg, A., Svedin, A., \& Vasil, G.~M. 2009, \href{http://dx.doi.org/10.1111/j.1365-2966.2009.14646.x}{\JournalTitle{MNRAS}, 395, 1599}

\bibitem[{Brown {et~al.}(1989)Brown, Byrne, \& Hindmarsh}]{vode}
Brown, P.~N., Byrne, G.~D., \& Hindmarsh, A.~C. 1989, \href{http://dx.doi.org/10.1137/0910062}{\JournalTitle{SIAM J. Sci. and Stat. Comput.}, 10, 1038}

\bibitem[{Bryan {et~al.}(1995)Bryan, Norman, Stone, Cen, \& Ostriker}]{bryan:1995}
Bryan, G.~L., Norman, M.~L., Stone, J.~M., Cen, R., \& Ostriker, J.~P. 1995, \href{http://dx.doi.org/10.1016/0010-4655(94)00191-4}{\JournalTitle{Comput. Phys. Commun.}, 89, 149}

\bibitem[{{Casanova} {et~al.}(2011{\natexlab{a}}){Casanova}, {Jos{\'e}}, {Garc{\'\i}a-Berro}, {Calder}, \& {Shore}}]{casanova:2011a}
{Casanova}, J., {Jos{\'e}}, J., {Garc{\'\i}a-Berro}, E., {Calder}, A., \& {Shore}, S.~N. 2011{\natexlab{a}}, \href{http://dx.doi.org/10.1051/0004-6361/201015895}{\JournalTitle{\aap}, 527, A5}

\bibitem[{{Casanova} {et~al.}(2011{\natexlab{b}}){Casanova}, {Jos{\'e}}, {Garc{\'\i}a-Berro}, {Shore}, \& {Calder}}]{casanova:2011b}
{Casanova}, J., {Jos{\'e}}, J., {Garc{\'\i}a-Berro}, E., {Shore}, S.~N., \& {Calder}, A.~C. 2011{\natexlab{b}}, \href{http://dx.doi.org/10.1038/nature10520}{\JournalTitle{\nat}, 478, 490}

\bibitem[{{Casanova} {et~al.}(2016){Casanova}, {José}, {García-Berro}, \& {Shore}}]{casanova:2016}
{Casanova}, J., {José}, J., {García-Berro}, E., \& {Shore}, S.~N. 2016, \href{http://dx.doi.org/10.1051/0004-6361/201628707}{\JournalTitle{A\&A}, 595, A28}

\bibitem[{{Casanova} {et~al.}(2018){Casanova}, {José}, \& {Shore}}]{casanova:2018}
{Casanova}, J., {José}, J., \& {Shore}, S.~N. 2018, \href{http://dx.doi.org/10.1051/0004-6361/201833422}{\JournalTitle{A\&A}, 619, A121}

\bibitem[{{Casanova} {et~al.}(2010){Casanova}, {José, J.}, {García-Berro, E.}, {Calder, A.}, \& {Shore, S. N.}}]{casanova:2010}
{Casanova}, J., {José, J.}, {García-Berro, E.}, {Calder, A.}, \& {Shore, S. N.} 2010, \href{http://dx.doi.org/10.1051/0004-6361/201014178}{\JournalTitle{A\&A}, 513, L5}

\bibitem[{{Cloutman} \& {Eoll}(1976)}]{cloutman:1976}
{Cloutman}, L.~D. \& {Eoll}, J.~G. 1976, \href{http://dx.doi.org/10.1086/154411}{\JournalTitle{\apj}, 206, 548}

\bibitem[{Colella(1990)}]{colella:1990}
Colella, P. 1990, \href{http://dx.doi.org/https://doi.org/10.1016/0021-9991(90)90233-Q}{\JournalTitle{J.~Comput.~Phys.}, 87, 171}

\bibitem[{Colella \& Woodward(1984)}]{ppm}
Colella, P. \& Woodward, P.~R. 1984, \href{http://dx.doi.org/10.1016/0021-9991(84)90143-8}{\JournalTitle{J. Comput. Phys.}, 54, 174}

\bibitem[{Cyburt {et~al.}(2010)Cyburt, Amthor, Ferguson, Meisel, Smith, Warren, Heger, Hoffman, Rauscher, Sakharuk, Schatz, Thielemann, \& Wiescher}]{cyburt:2010}
Cyburt, R.~H., Amthor, A.~M., Ferguson, R., {et~al.} 2010, \href{http://dx.doi.org/10.1088/0067-0049/189/1/240}{\JournalTitle{ApJS}, 189, 240}

\bibitem[{Eggleton(1971)}]{eggleton:1971}
Eggleton, P.~P. 1971, \href{http://dx.doi.org/10.1093/mnras/151.3.351}{\JournalTitle{MNRAS}, 151, 351}

\bibitem[{{Eiden} {et~al.}(2020){Eiden}, {Zingale}, {Harpole}, {Willcox}, {Cavecchi}, \& {Katz}}]{eiden:2020}
{Eiden}, K., {Zingale}, M., {Harpole}, A., {et~al.} 2020, \href{http://dx.doi.org/10.3847/1538-4357/ab80bc}{\JournalTitle{\apj}, 894, 6}

\bibitem[{{Fan} {et~al.}(2019){Fan}, {Nonaka}, {Almgren}, {Harpole}, \& {Zingale}}]{maestroex}
{Fan}, D., {Nonaka}, A., {Almgren}, A.~S., {Harpole}, A., \& {Zingale}, M. 2019, \href{http://dx.doi.org/10.3847/1538-4357/ab4f75}{\JournalTitle{\apj}, 887, 212}

\bibitem[{{Fryxell} {et~al.}(1989){Fryxell}, {M{\"u}ller}, \& {Arnett}}]{prometheus}
{Fryxell}, B.~A., {M{\"u}ller}, E., \& {Arnett}, D. 1989, \JournalTitle{MPA Preprint 449}

\bibitem[{{Fryxell} \& {Woosley}(1982)}]{fryxell:1982}
{Fryxell}, B.~A. \& {Woosley}, S.~E. 1982, \href{http://dx.doi.org/10.1086/160344}{\JournalTitle{\apj}, 261, 332}

\bibitem[{{Fujimoto}(1988)}]{fujimoto:1988}
{Fujimoto}, M.~Y. 1988, \JournalTitle{\aap}, 198, 163

\bibitem[{Gehrz {et~al.}(1998)Gehrz, Truran, Williams, \& Starrfield}]{gehrz:1998}
Gehrz, R.~D., Truran, J.~W., Williams, R.~E., \& Starrfield, S. 1998, \href{http://dx.doi.org/10.1086/316107}{\JournalTitle{Publications of the Astronomical Society of the Pacific}, 110, 3}

\bibitem[{{Glasner} \& {Livne}(1995)}]{glasner:1995}
{Glasner}, S.~A. \& {Livne}, E. 1995, \href{http://dx.doi.org/10.1086/187911}{\JournalTitle{\apjl}, 445, L149}

\bibitem[{Glasner {et~al.}(1997)Glasner, Livne, \& Truran}]{glasner:1997}
Glasner, S.~A., Livne, E., \& Truran, J.~W. 1997, \href{http://dx.doi.org/10.1086/303561}{\JournalTitle{ApJ}, 475, 754}

\bibitem[{Glasner {et~al.}(2005)Glasner, Livne, \& Truran}]{glasner:2005}
Glasner, S.~A., Livne, E., \& Truran, J.~W. 2005, \href{http://dx.doi.org/10.1086/429482}{\JournalTitle{ApJ}, 625, 347}

\bibitem[{Glasner {et~al.}(2007)Glasner, Livne, \& Truran}]{glasner:2007}
Glasner, S.~A., Livne, E., \& Truran, J.~W. 2007, \JournalTitle{ApJ}, 665, 1321

\bibitem[{{Iben} \& {MacDonald}(1985)}]{iben:1985}
{Iben}, I., J. \& {MacDonald}, J. 1985, \href{http://dx.doi.org/10.1086/163473}{\JournalTitle{\apj}, 296, 540}

\bibitem[{Iben~Jr {et~al.}(1992)Iben~Jr, Fujimoto, \& MacDonald}]{iben:1992}
Iben~Jr, I., Fujimoto, M.~Y., \& MacDonald, J. 1992, \JournalTitle{ApJ}, 388, 521

\bibitem[{{José}(2024)}]{jordi:2024}
{José}, J. 2024, \href{http://dx.doi.org/10.1051/epjconf/202429701006}{\JournalTitle{EPJ Web Conf.}, 297, 01006}

\bibitem[{{José} {et~al.}(2020){José}, {Shore}, \& {Casanova}}]{casanova:2020}
{José}, J., {Shore}, S.~N., \& {Casanova}, J. 2020, \href{http://dx.doi.org/10.1051/0004-6361/201936893}{\JournalTitle{A\&A}, 634, A5}

\bibitem[{{K{\"a}ppeli} \& {Mishra}(2016)}]{kappeli:2016}
{K{\"a}ppeli}, R. \& {Mishra}, S. 2016, \href{http://dx.doi.org/10.1051/0004-6361/201527815}{\JournalTitle{\aap}, 587, A94}

\bibitem[{Katz {et~al.}(2016)Katz, Zingale, Calder, Swesty, Almgren, \& Zhang}]{wdmergerI}
Katz, M.~P., Zingale, M., Calder, A.~C., {et~al.} 2016, \href{http://dx.doi.org/10.3847/0004-637x/819/2/94}{\JournalTitle{ApJ}, 819, 94}

\bibitem[{Katz {et~al.}(2020)Katz, Almgren, Sazo, Eiden, Gott, Harpole, Sexton, Willcox, Zhang, \& Zingale}]{castro-gpu}
Katz, M.~P., Almgren, A., Sazo, M.~B., {et~al.} 2020, in Proceedings of the International Conference for High Performance Computing, Networking, Storage and Analysis, SC '20 (IEEE Press)

\bibitem[{{Kercek} {et~al.}(1998){Kercek}, {Hillebrandt}, \& {Truran}}]{kercek:1998}
{Kercek}, A., {Hillebrandt}, W., \& {Truran}, J.~W. 1998, \href{http://dx.doi.org/10.48550/arXiv.astro-ph/9801054}{\JournalTitle{\aap}, 337, 379}

\bibitem[{{Kercek} {et~al.}(1999){Kercek}, {Hillebrandt}, \& {Truran}}]{kercek:1999}
{Kercek}, A., {Hillebrandt}, W., \& {Truran}, J.~W. 1999, \href{http://dx.doi.org/10.48550/arXiv.astro-ph/9811259}{\JournalTitle{\aap}, 345, 831}

\bibitem[{{Kippenhahn} \& {Thomas}(1978)}]{kippenhahn:1978}
{Kippenhahn}, R. \& {Thomas}, H.~C. 1978, \JournalTitle{\aap}, 63, 265

\bibitem[{{Kovetz} \& {Prialnik}(1985)}]{prialnik:1985}
{Kovetz}, A. \& {Prialnik}, D. 1985, \href{http://dx.doi.org/10.1086/163117}{\JournalTitle{\apj}, 291, 812}

\bibitem[{{Kutter} \& {Sparks}(1972)}]{kutter:1972}
{Kutter}, G.~S. \& {Sparks}, W.~M. 1972, \href{http://dx.doi.org/10.1086/151567}{\JournalTitle{\apj}, 175, 407}

\bibitem[{{Kutter} \& {Sparks}(1980)}]{kutter:1980}
{Kutter}, G.~S. \& {Sparks}, W.~M. 1980, \href{http://dx.doi.org/10.1086/158187}{\JournalTitle{\apj}, 239, 988}

\bibitem[{{Kutter} \& {Sparks}(1987)}]{kutter:1987}
{Kutter}, G.~S. \& {Sparks}, W.~M. 1987, \href{http://dx.doi.org/10.1086/165637}{\JournalTitle{\apj}, 321, 386}

\bibitem[{{Kutter} \& {Sparks}(1989)}]{kutter:1989}
{Kutter}, G.~S. \& {Sparks}, W.~M. 1989, \href{http://dx.doi.org/10.1086/167452}{\JournalTitle{\apj}, 340, 985}

\bibitem[{Livio \& Truran(1990)}]{livio:1990}
Livio, M. \& Truran, J.~W. 1990, \href{http://dx.doi.org/https://doi.org/10.1111/j.1749-6632.1990.tb37801.x}{\JournalTitle{Annals of the New York Academy of Sciences}, 617, 126}

\bibitem[{{Livio} \& {Truran}(1994)}]{livio:1994}
{Livio}, M. \& {Truran}, J.~W. 1994, \href{http://dx.doi.org/10.1086/174024}{\JournalTitle{\apj}, 425, 797}

\bibitem[{{Livne}(1993)}]{livne:1993}
{Livne}, E. 1993, \href{http://dx.doi.org/10.1086/172950}{\JournalTitle{\apj}, 412, 634}

\bibitem[{{MacDonald}(1983)}]{mcdonald:1983}
{MacDonald}, J. 1983, \href{http://dx.doi.org/10.1086/161368}{\JournalTitle{\apj}, 273, 289}

\bibitem[{Miller \& Colella(2002)}]{millercolella:2002}
Miller, G. \& Colella, P. 2002, \href{http://dx.doi.org/10.1006/jcph.2002.7158}{\JournalTitle{J. Comput. Phys.}, 183, 26}

\bibitem[{Minoshima \& Miyoshi(2021)}]{hlld}
Minoshima, T. \& Miyoshi, T. 2021, \href{http://dx.doi.org/https://doi.org/10.1016/j.jcp.2021.110639}{\JournalTitle{Journal of Computational Physics}, 446, 110639}

\bibitem[{Orio \& Shaviv(1993)}]{orio:1993}
Orio, M. \& Shaviv, G. 1993, \href{http://dx.doi.org/10.1007/BF00626882}{\JournalTitle{Astrophysics and Space Science}, 202, 273}

\bibitem[{Ouellette(2012)}]{ouellette:2012}
Ouellette, N.~T. 2012, \href{http://dx.doi.org/10.1063/pt.3.1570}{\JournalTitle{Phys. Today}, 65, 68}

\bibitem[{Parete-Koon {et~al.}(2003)Parete-Koon, Hix, Smith, Starrfield, Bardayan, Guidry, \& Mezzacappa}]{koon:2003}
Parete-Koon, S., Hix, W.~R., Smith, M.~S., {et~al.} 2003, \href{http://dx.doi.org/10.1086/378979}{\JournalTitle{ApJ}, 598, 1239}

\bibitem[{Prialnik \& Kovetz(1984)}]{prialnik:1984}
Prialnik, D. \& Kovetz, A. 1984, \href{http://dx.doi.org/10.1086/162107}{\JournalTitle{ApJ}, 281, 367}

\bibitem[{{Prialnik} {et~al.}(1979){Prialnik}, {Shara}, \& {Shaviv}}]{prialnik:1979}
{Prialnik}, D., {Shara}, M.~M., \& {Shaviv}, G. 1979, \JournalTitle{\aap}, 72, 192

\bibitem[{{Shankar} \& {Arnett}(1994)}]{shankar:1994}
{Shankar}, A. \& {Arnett}, D. 1994, \href{http://dx.doi.org/10.1086/174637}{\JournalTitle{\apj}, 433, 216}

\bibitem[{{Shankar} {et~al.}(1992){Shankar}, {Arnett}, \& {Fryxell}}]{shankar:1992}
{Shankar}, A., {Arnett}, D., \& {Fryxell}, B.~A. 1992, \href{http://dx.doi.org/10.1086/186461}{\JournalTitle{\apjl}, 394, L13}

\bibitem[{Shara(1981)}]{shara:1981}
Shara, M.~M. 1981, \href{http://dx.doi.org/10.1086/158657}{\JournalTitle{\apj}, 243, 926}

\bibitem[{Shara(1982)}]{shara:1982}
Shara, M.~M. 1982, \href{http://dx.doi.org/10.1086/160376}{\JournalTitle{ApJ}, 261, 649}

\bibitem[{Smith {et~al.}(2023)Smith, Johnson, Chen, Eiden, Willcox, Boyd, Cao, DeGrendele, \& Zingale}]{pynucastro2}
Smith, A.~I., Johnson, E.~T., Chen, Z., {et~al.} 2023, \href{http://dx.doi.org/10.3847/1538-4357/acbaff}{\JournalTitle{ApJ}, 947, 65}

\bibitem[{Smith~Clark \& Zingale(2025)}]{smith:2025}
Smith~Clark, A. \& Zingale, M. 2025, \href{http://dx.doi.org/10.5281/zenodo.14939250}{\JournalTitle{Zenodo record 10.5281/zenodo.14939250}}

\bibitem[{{Sparks} {et~al.}(1976){Sparks}, {Starrfield}, \& {Truran}}]{sparks:1978}
{Sparks}, W.~M., {Starrfield}, S., \& {Truran}, J.~W. 1976, \href{http://dx.doi.org/10.1086/154668}{\JournalTitle{\apj}, 208, 819}

\bibitem[{Strang(1968)}]{strang:1968}
Strang, G. 1968, \href{http://dx.doi.org/10.1137/0705041}{\JournalTitle{SIAM J. Numer. Anal.}, 5, 506}

\bibitem[{Timmes(2000)}]{timmes:2000}
Timmes, F.~X. 2000, \href{http://dx.doi.org/10.1086/308203}{\JournalTitle{ApJ}, 528, 913}

\bibitem[{Timmes \& Arnett(1999)}]{timmes:1999}
Timmes, F.~X. \& Arnett, D. 1999, \href{http://dx.doi.org/10.1086/313271}{\JournalTitle{ApJS}, 125, 277}

\bibitem[{Timmes \& Swesty(2000)}]{timmes_swesty:2000}
Timmes, F.~X. \& Swesty, F.~D. 2000, \href{http://dx.doi.org/10.1086/313304}{\JournalTitle{ApJS}, 126, 501}, source code obtained from http://cococubed.asu.edu/codes/eos.shtml/helmholtz.tbz

\bibitem[{Wiescher {et~al.}(1999)Wiescher, Görres, \& Schatz}]{wiescher:1999}
Wiescher, M., Görres, J., \& Schatz, H. 1999, \href{http://dx.doi.org/10.1088/0954-3899/25/6/201}{\JournalTitle{Journal of Physics G: Nuclear and Particle Physics}, 25, R133}

\bibitem[{Zhang {et~al.}(2019)Zhang, Almgren, Beckner, Bell, Blaschke, Chan, Day, Friesen, Gott, Graves, Katz, Myers, Nguyen, Nonaka, Rosso, Williams, \& Zingale}]{amrex_joss}
Zhang, W., Almgren, A., Beckner, V., {et~al.} 2019, \href{http://dx.doi.org/10.21105/joss.01370}{\JournalTitle{JOSS}, 4, 1370}

\bibitem[{Zingale(2024)}]{hse-rnaas}
Zingale, M. 2024, \href{http://dx.doi.org/10.3847/2515-5172/ad76b0}{\JournalTitle{Research Notes of the AAS}, 8, 219}

\bibitem[{Zingale {et~al.}(2024)Zingale, Chen, Johnson, Katz, \& Smith~Clark}]{castro-massive-stars}
Zingale, M., Chen, Z., Johnson, E.~T., Katz, M.~P., \& Smith~Clark, A. 2024, \href{http://dx.doi.org/10.3847/1538-4357/ad8a66}{\JournalTitle{ApJ}, 977, 30}

\bibitem[{Zingale {et~al.}(2021)Zingale, Katz, Willcox, \& Harpole}]{strang_rnaas}
Zingale, M., Katz, M.~P., Willcox, D.~E., \& Harpole, A. 2021, \href{http://dx.doi.org/10.3847/2515-5172/abf3cb}{\JournalTitle{Research Notes of the AAS}, 5, 71}

\bibitem[{Zingale {et~al.}(2002)Zingale, Dursi, ZuHone, Calder, Fryxell, Plewa, Truran, Caceres, Olson, Ricker, Riley, Rosner, Siegel, Timmes, \& Vladimirova}]{zingale:2002}
Zingale, M., Dursi, L.~J., ZuHone, J., {et~al.} 2002, \href{http://dx.doi.org/10.1086/342754}{\JournalTitle{ApJS}, 143, 539}

\end{thebibliography}

\end{document}